\documentclass[a4paper,11pt]{article}
\pdfoutput=1 

\usepackage{amsmath}
\usepackage{amssymb}
\usepackage{mathtools}
\usepackage{stackengine,scalerel}
\usepackage{braket}
\usepackage{jheppub} 

\usepackage[T1]{fontenc} 

\newcommand\textaltcolon{\ensurestackMath{\stackon[0.3ex]{\circ}{\circ}}}
\newcommand\altcolon{\savestack\Tmp{\raisebox{-2.7pt}{$\textaltcolon$}}%
  \dp\Tmpcontent=\dimexpr\dp\Tmpcontent-2.7pt\relax%
  \mathrel{\scalerel*{\Tmp}{:}}}

\title{\boldmath The free field realisation of the BVW string}

\author[]{Matthias R.\ Gaberdiel,}
\author[]{Kiarash Naderi}
\author[]{and Vit Sriprachyakul}

\affiliation[]{Institute for Theoretical Physics, ETH Zurich \\ 8093 Zurich, Switzerland}

\emailAdd{gaberdiel@itp.phys.ethz.ch}
\emailAdd{knaderi@phys.ethz.ch}
\emailAdd{vsriprachyak@phys.ethz.ch}

\abstract{The symmetric orbifold of $\mathbb{T}^4$ was recently shown to be exactly dual to string theory on ${\rm AdS}_3\times {\rm S}^3 \times \mathbb{T}^4$ with minimal ($k=1$) NS-NS flux. The worldsheet theory is best formulated in terms of the hybrid formalism of Berkovits, Vafa \& Witten (BVW), in terms of which the ${\rm AdS}_3\times {\rm S}^3$ factor is described by a $\mathfrak{psu}(1,1|2)_k$ WZW model. 
At level $k=1$, $\mathfrak{psu}(1,1|2)_1$ has a free field realisation that is obtained from that of $\mathfrak{u}(1,1|2)_1$ upon setting a $\mathfrak{u}(1)$ field, often called $Z$, to zero. We show that the free field version of the ${\cal N}=2$ generators of BVW (whose cohomology defines the physical states) does not give rise to an ${\cal N}=2$ algebra, but is rather contaminated by terms proportional to the $Z$-field. We also show how to overcome this problem by introducing additional ghost fields that implement the quotienting by $Z$.}

\numberwithin{equation}{section}

\def\be{\begin{equation}}
\def\ee{\end{equation}}

\begin{document}
\maketitle
\flushbottom

\section{Introduction}

The string theory background dual to the actual symmetric orbifold of $\mathbb{T}^4$ has recently been identified in \cite{Eberhardt:2018ouy,Eberhardt:2019ywk}, see also \cite{Gaberdiel:2018rqv,Giribet:2018ada} for earlier work. The string background has pure (and minimal, i.e.\ $k=1$) NS-NS flux, and can therefore be described be an exactly solvable worldsheet WZW model. Given the familiar problems with the $k=1$ theory in the usual R-NS approach of \cite{Maldacena:2000hw}, the best description utilises the so-called hybrid formalism of Berkovits, Vafa and Witten (BVW) \cite{BVW}. Then the relevant WZW model is based on the superalgebra $\mathfrak{psu}(1,1|2)_k$, for which the case $k=1$ is unproblematic. From this viewpoint the breakdown of the naive R-NS description at $k=1$ manifests itself in that  $\mathfrak{psu}(1,1|2)_k$ only possesses short representations at level $k=1$ \cite{Eberhardt:2018ouy}.

The full worldsheet theory of the BVW hybrid string consists of a $\mathfrak{psu}(1,1|2)_k$ WZW model, together with a topologically twisted $\mathbb{T}^4$ theory. This is to say, the four free bosons of $\mathbb{T}^4$ are conventional free fields (and their derivatives define fields of conformal dimension $h=1$), while of the four free fermions, two have conformal dimension $h=1$, while the other two have $h=0$. In addition, the theory has a number of ghost fields that arise from the original R-NS description by a sequence of field redefinitions \cite{BVW}. The worldsheet theory has an ${\cal N}=2$ superconformal symmetry \cite{BV}, which can be enhanced to ${\cal N}=4$  following \cite{Berkovits:1994vy}, and the physical spectrum can be characterised as a certain double cohomology of this (topological) ${\cal N}=4$ string \cite{BVW}. (Alternatively, one may describe the physical states in terms of the critical ${\cal N}=2$ theory.) While this is an elegant description of the spectrum, it is actually quite complicated to determine this cohomology from first principles, and only rather partial results are known to date \cite{Troost:2011fd,Gaberdiel:2011vf,Seb}. 
\smallskip

Just like the $\mathfrak{su}(2)_k$ algebra, also $\mathfrak{psu}(1,1|2)_k$ has a free field realisation at level $k=1$ \cite{Lesage:2002ch,Ridout:2010jk,Eberhardt:2018ouy}, and not surprisingly this is the most efficient  descripition of the theory. For example, while the localisation property of the worldsheet correlators could already be seen using the Ward identities of the $\mathfrak{sl}(2,\mathbb{R})$ affine algebra \cite{Eberhardt:2019ywk},\footnote{This analysis was later generalised to higher genus \cite{Eberhardt:2020akk}, as well as to the case of $k>1$ \cite{Dei:2021xgh,Dei:2021yom} where the generic solution does not localise.} the proof that the localised solution is in fact the only solution was only achieved using the free field description \cite{DGGK} of the BVW formalism. (The higher genus version of this analysis was later done in \cite{Knighton:2020kuh}.)

In view of this situation it is therefore important to understand the free field realisation of the $k=1$ BVW theory in detail, in particular with a view towards deriving its cohomology from first principles. On the face of it, this sounds like a rather straightforward exercise, but the situation is actually somewhat more complicated. The  free field realisation of $\mathfrak{psu}(1,1|2)_1$ is very much like the free field realisation of $\mathfrak{su}(2)_1$ in terms of four free fermions: it actually leads to $\mathfrak{u}(1,1|2)_1$ (or $\mathfrak{u}(2)_1$, for the case of $\mathfrak{su}(2)_1$), and in order to obtain $\mathfrak{psu}(1,1|2)_1$ one needs to set to zero a $\mathfrak{u}(1)$ field that is usually denoted by $Z$. Usually, this does not cause any significant problems, but in the present case the situation is more subtle since the construction of the ${\cal N}=2$ fields depends crucially, for example, on the central charge of the $\mathfrak{psu}(1,1|2)_1$ factor (which is $c=-2$), while the free field realisation itself has $c=0$. As a consequence (and as we explain in this paper), there does not seem to be any simple way of defining the ${\cal N}=2$ (or indeed the ${\cal N}=4$) fields in the free theory itself; instead the natural ${\cal N}=2$ fields have correction terms in their OPEs that are proportional to the $\mathfrak{u}(1)$ field $Z$ that needs to be set to zero in order to obtain $\mathfrak{psu}(1,1|2)$ --- see eqs.~(\ref{mis1}) and (\ref{mis2}) below.

Obviously, in order to define the cohomology as in \cite{BVW}, it is important that the OPEs are exactly obeyed; for example, the fact that the $G^+ G^+$ OPE is trivial implies that the corresponding zero mode $G^+_0$ is nilpotent (and can hence play the role of one of the BRST operators in the topological ${\cal N}=4$ string). Given that the correction term involves the $Z$ field one may expect that it should be possible to restore the ${\cal N}=2$ structure if, {\em at the same time}, also the quotienting out of the $Z$ field is implemented. Since the latter is effectively a coset construction, it can be described, in a standard BRST formalism, by introducing additional ghost fields \cite{Karabali:1989dk}. Our setup is slightly different --- the physical states are those of the ${\cal N}=4$ topological string, and not just some standard BRST cohomology --- and so we cannot literally use the construction of \cite{Karabali:1989dk}. However, it is possible to imitate aspects of their construction: after introducing the same set of ghost fields as in \cite{Karabali:1989dk}, we can modify the ${\cal N}=2$ (and ${\cal N}=4$) generators so that they form an actual ${\cal N}=2$ (and ${\cal N}=4$) algebra. Furthermore, we can argue that the corresponding ${\cal N}=4$ cohomology will implement both the physical state cohomology of \cite{BVW}, as well as remove the unwanted $\mathfrak{u}(1)$ field $Z$. 
\medskip

The paper is organised as follows. In Section~\ref{sec:BVW} we give a brief review of the BVW construction for the ${\cal N}=2$ algebra; this follows largely their work, but we spell out some of the details of the relevant calculations that play an important role in the following. We also discuss the construction of the global $\mathfrak{psu}(1,1|2)$ charges in Section~\ref{sec:global}. In Section~\ref{FF} we make a natural ansatz for the ${\cal N}=2$ generators in terms of the free fields, and explain in detail which OPEs get modified, see eqs.~(\ref{mis1}) and (\ref{mis2}). We also discuss the fate of the global symmetry generators in Section~\ref{sec:global-free}. In  Section~\ref{sec:n=4-modified} we add the additional ghost fields and find the actual ${\cal N}=2$ (and ${\cal N}=4$) algebra generators. We also study its cohomology in Section~\ref{sec:modifiedcohomology}, and discuss how the global $\mathfrak{psu}(1,1|2)$ charges work in this modified setting. Our conclusions are summarised in Section~\ref{sec:concl}, and there are a number of appendices where some of our conventions and the technical details of our analysis are spelled out.

\section{A review of the BVW construction for the $\mathcal{N}=2$ algebra}\label{sec:BVW}

In the hybrid formulation of Berkovits, Vafa and Witten \cite{BVW} the physical states are characterised by a certain double coholomogy whose BRST operators come from an ${\cal N}=4$ superconformal algebra. (Alternatively, one can characterise the physical states as those of the critical ${\cal N}=2$ theory \cite{BV,Berkovits:1994vy}.) In this section we briefly review the construction of this ${\cal N}=2$ algebra --- the extension to the ${\cal N}=4$ algebra is then relatively straightforward. Following \cite{BVW} we shall begin by considering the flat space situation, and then generalise to ${\rm AdS}_3 \times {\rm S}^3$. We shall also discuss how the global supersymmetries of the string background are realised on the worldsheet.

\subsection{Flat space}\label{sec:flat}

Let us begin by reviewing the $\mathcal{N}=2$ algebra of \cite{BVW}  for the case of a flat space background, i.e.\ for $\mathbb{R}^{6} \times \mathbb{T}^4$.  According to Section~4 of \cite{BVW}, the ${\cal N}=2$ generators take the form 
{\allowdisplaybreaks
\begin{eqnarray}
T  & = & T_{\rm flat} - \frac{1}{2} \partial \rho \partial \rho - \frac{1}{2} \partial \sigma \partial \sigma 
+ \frac{3}{2} \partial^2( \rho+i\sigma) + T_{C} \\
G^+ & = &  \underbrace{- e^{-2\rho - i\sigma} (p)^4}_{Q_2} 
- \underbrace{ \frac{i}{2} e^{-\rho} p_a p_b \partial x^{ab}}_{Q_1} + \\
& & \qquad 
+ \underbrace{e^{i\sigma} \Bigl( T_{\rm flat} - \frac{1}{2} \partial (\rho + i \sigma) \partial (\rho + i \sigma) + \frac{1}{2} \partial^2 (\rho + i \sigma) \Bigr) }_{Q_0}
+ G^+_C \nonumber\\
G^- & = & 2 e^{-i\sigma} + G^-_{C} \\
J & = & \partial (\rho + i \sigma) + J_C \ , 
\label{N2genflat}
\end{eqnarray}}

\vspace*{-0.5cm}
\noindent  as follows upon applying a similarity transformation to the generators of \cite{BV}. Here $x^{m}$, and $p_{a}$, $\theta^a$ are the flat-space fields, with the bosonic $x^m$ fields satisfying
\be 
x^m(z)x^n(w) \sim -\eta^{mn} \ln(z-w)\ ,
\ee
where $\eta^{mn}$ is the $6d$ Euclidean metric, and $m$ runs from $1,\ldots,6$. The fermionic degrees of freedom are described in a first-order formalism, and we have 
\be
p_a(z)\theta^b(w) \sim \dfrac{\delta^b_a}{z-w} \ ,  
\ee
where $a,b\in \{1,2,3,4\}$, and $(p,\theta)$ have conformal dimensions $(1,0)$. The corresponding flat space stress tensor $T_{\rm flat}$ is then given by 
\be\label{Tint0}
T_{\rm flat}=- \frac{1}{2} \partial x^m \partial x_m - \, p_a \partial \theta^a \ .
\ee
In addition, we have the twisted ${\cal N}=2$ fields $T_C$, $G^\pm_C$ and $J_C$, associated to the $\mathbb{T}^4$ factor --- we will sometimes refer to them as the `compact' generators --- while $\rho$ and $\sigma$  are ghost fields satisfying 
\be
\rho(z)\, \rho(w) \sim \sigma(z)\, \sigma(w) \sim -\ln(z-w) \ .
\ee
Their central charges are $c(\rho)=28$ and $c(\sigma)=-26$, and $\sigma$ is the bosonisation of the usual $(b,c)$ diffeomorphism ghosts. 
We should mention that relative to \cite{BVW},  we have changed the sign of the $Q_{1}$ term; this sign is irrelevant for the ${\cal N}=2$ algebra, but is more convenient to ensure that the generators commute with the global symmetry generators, see below. We have also changed some other factors and signs relative to \cite{BVW} so that these generators conform with our conventions for the $\mathcal{N}=2$ algebra, see Appendix~\ref{N2}. Finally, all of the above expressions are suitably normal ordered, and we shall discuss the precise definition of the normal ordering in more detail for the curved background below. 



It is in principle straightforward to check that these generators indeed satisfy a twisted $\mathcal{N}=2$ algebra with $c=6$, but the calculation is a bit tedious.\footnote{Note that the Virasoro algebra of this twisted ${\cal N}=2$ algebra does not have a central term, but the central charge parameter $c$ appears for example in the $[L_m,J_n]$ commutator, see eqs.~(\ref{N2twis}).}  Since we are about to discuss the generalisation to the curved set-up (where we replace the flat-space fields by the generators of $\mathfrak{psu}(1,1|2)_k$), let us highlight a few key steps and phrase them in ways that will generalise. First of all, the most non-trivial relation to check is that the $G^{+}G^{+}$ OPE is regular. Since the different terms in $G^+$ come with different exponentials of the ghost fields, this is equivalent to requiring that we have separately 
\begin{align}
Q_{2}Q_{0}+Q_{1}Q_{1}+Q_{0}Q_{2} & \sim  0 \ , \label{Q2Q0} \\ 
Q_{1}Q_{0}+Q_{0}Q_{1} & \sim 0 \ ,  \label{Q1Q0} \\
Q_{0}Q_{0} & \sim 0 \ .  \label{Q0Q0}
\end{align}
The other combinations are easily seen to be regular; for example,  the $Q_{2}Q_{2}$ OPE is regular since the $p$'s do not have contractions among themselves, while the contraction of the exponentials gives
\be\label{2.11}
\exp\Bigl(-2\rho(z)-i\sigma(z)-2\rho(w)-i\sigma(w) \Bigr) \, (p)^4(z)(p)^4(w)\, (z-w)^{-3}\ ,
\ee
which, by expanding to second order in $(z-w)$, is regular. (Here $(p)^4=p_1p_2p_3p_4$.) The other terms work similarly.

Returning to the above terms, we note that the $Q_{0}Q_{0}$ OPE, see eq.~(\ref{Q0Q0}), is regular provided that $T_{\rm flat}$ has central charge $c=-2$, see Appendix~\ref{GG} for more details. This is the case, as follows from (\ref{Tint0}). 

Next we consider the $Q_{1}Q_{0}+Q_{0}Q_{1}$ term, see eq.~(\ref{Q1Q0}). If we write $Q_{1}=e^{-\rho}R$, where $R$ does not depend on the ghost fields, one can show that eq.~(\ref{Q1Q0}) is regular provided that $R$ is a primary field of weight $3$ with respect to $T_{\rm flat}$. For the situation at hand this is the case. 


Unfortunately, for the $Q_{2}Q_{0}+Q_{1}Q_{1}+Q_{0}Q_{2}$ term, see eq.~(\ref{Q2Q0}), there is no generic simplification, and we need to work it out explicitly. This is straightforward to do in the flat space case, and the complete expression is indeed regular.

%

\subsection{Generalisation to ${\rm AdS}_3 \times {\rm S}^3$}\label{curve}

Next we want to replace the $\mathbb{R}^6$ fields by the generators of $\mathfrak{psu}(1,1|2)_k$. Following \cite{BVW} the corresponding ${\cal N}=2$ generators are then 
\begin{subequations}\label{N2gencur}
\begin{align}
T  & =  T_{\rm int} - \frac{1}{2} \partial \rho \partial \rho - \frac{1}{2} \partial \sigma \partial \sigma 
+ \frac{3}{2} \partial^2( \rho+i\sigma) + T_{C}  \label{Tdef} \\
G^+ & =   \underbrace{- e^{-2\rho - i\sigma} (S)^4}_{Q_2} 
+ \underbrace{ e^{-\rho} R}_{Q_1} 
+ \underbrace{e^{i\sigma} \Bigl( T_{\rm int} - \frac{1}{2} \partial (\rho + i \sigma) \partial (\rho + i \sigma) + \frac{1}{2} \partial^2 (\rho + i \sigma) \Bigr) }_{Q_0}
+ G^+_C \label{GPdef}   \\ 
G^- & =  2 e^{-i\sigma} + G^-_{C} \\
J & =  \partial (\rho + i \sigma) + J_C \ , \label{eq:j-curved}
\end{align}
\end{subequations}
where now 
\begin{eqnarray}
T_{\rm int}&=&\dfrac{1}{8k}\, \epsilon^{abcd}K^{ab}K^{cd}-\dfrac{1}{k}S^{a}_{1}S^{a}_{2} 
\label{curvegen}\\
(S)^{4}&=&\dfrac{1}{24}\, \epsilon^{abcd} \, S^{a}_{1}S^{b}_{1}S^{c}_{1}S^{d}_{1}=S^{1}_{1}S^{2}_{1}S^{3}_{1}S^{4}_{1}  \label{S4def} \\
R&=& -\dfrac{1}{2\sqrt{k}}\Bigl( K^{ab}S^{a}_{1}S^{b}_{1}+4S^{a}_{1} \partial S^{a}_{1} \Bigr) \ . \label{Rdef}
\end{eqnarray}
Here $K^{ab}$ are the bosonic and $S^{a}_{\alpha}$ the fermionic generators of $\mathfrak{psu}(1,1|2)_k$, see Appendix~\ref{psulevk} for our conventions. All of these expressions are normal ordered with the prescription that 
\begin{equation}
\begin{aligned}
: A_m B_n: = \left\{ \begin{array}{ll}    ~~A_m B_n & \hbox{if $m \leq -1$} \\
                                                       \pm B_n A_m &  \hbox{if $m \geq 0$ ,}
                   \end{array} \right.
\label{normalord}
\end{aligned}
\end{equation}
and the sign depends on whether at least one of the currents $A$ and $B$ is bosonic,  or whether both are fermionic. We should mention that this does not uniquely specify the normal ordering of the $R$ term (since it involves more than two generators), and the correct prescription is described in eqs.~(\ref{A.17}) and (\ref{A.18}).\footnote{The normal ordering of the $(S)^4$ term is unproblematic because of the anti-symmetry of the $\epsilon$ tensor and the fact that the $S$ fields anti-commute.}

Given our comments above, it is relatively straightforward to check that this realises indeed the ${\cal N}=2$ algebra. In particular, $T_{\rm int}$ has central charge $c=-2$, and hence (\ref{Q0Q0}) is regular, while (\ref{Q1Q0}) is regular because $R$, defined in eq.~(\ref{Rdef}), is indeed primary with respect to $T_{\rm int}$ of conformal weight $3$. We have also checked (with some effort) that the $Q_{2}Q_{0}+Q_{1}Q_{1}+Q_{0}Q_{2}$ term in (\ref{Q2Q0}) is regular.

\subsection{Generators of the global symmetry}\label{sec:global}

Since the ${\cal N}=2$ generators define effectively the BRST cohomology of \cite{BVW}, one should expect that they commute with the global $\mathfrak{psu}(1,1|2)$ symmetries of the background. For the bosonic generators of $\mathfrak{psu}(1,1|2)$ this is essentially obvious since all the generators in (\ref{Tdef}) -- (\ref{eq:j-curved}) are singlets with respect to the $\mathfrak{sl}(2,\mathbb{R}) \oplus \mathfrak{su}(2)$ subalgebra of $\mathfrak{psu}(1,1|2)$, i.e.\ 
\be
{}[K^{ab}_0,T] = [K^{ab}_0,G^\pm] = [K^{ab}_0,J] = 0 \ . 
\ee
The situation is a bit more complicated for the fermionic generators of $\mathfrak{psu}(1,1|2)$. While we have 
\be\label{S1comm}
{}[(S^{a}_{1})_0,T] = \{(S^{a}_{1})_0,G^\pm\} = [(S^{a}_{1})_0,J] = 0 \ ,
\ee
the zero modes $(S^{a}_{2})_0$ do not directly anti-commute with $G^+$, as was also already noted in \cite{BVW}. Before we explain how to correct for this, let us mention that even the identity $\{(S^{a}_{1})_0,G^+\} = 0$ in (\ref{S1comm}) is not entirely obvious: while the OPE of $S^{l}_{1}$ with $Q_{2}$  is regular (and hence $(S^{a}_{1})_0$ anticommutes with the corresponding terms in $G^+$),\footnote{Note that both $e^{-\rho}$ and $e^{\pm i\sigma}$ are fermionic operators.} the situation for $Q_1$ is somewhat tricky and hinges on the correct normal ordering prescription for the $R$ term, see eqs.~(\ref{A.17}) and (\ref{A.18}). Finally, the OPE of $S^{a}_{1}$ with $T_{\rm int}$ has a double pole, which however does not contribute after performing the contour integral to extract the zero mode $(S^{a}_{1})_0$. 

Since the zero modes $(S^{a}_{2})_0$ do not directly anticommute with $G^+$ --- there are non-trivial contributions both from the $Q_2$ and the $Q_1$ terms of $G^+$ --- we need to modify them, and the correct prescription, following \cite{BVW}, is to add the term
\be\label{fermsymmod}
(\tilde{S}^a_2)_0 = (S^{a}_{2})_0 + \sqrt{k}\, \oint dz \, e^{-\rho-i\sigma}\, S^a_1(z) \ . 
\ee
The simple pole in the $Q_2 \cdot S^{a}_{2}$ OPE is now cancelled by that in the OPE of $Q_1$ with the correction term $e^{-\rho-i\sigma}\, S^a_1$, while the simple pole in the $Q_1 \cdot S^{a}_{2}$ OPE cancels against that of the $Q_0 \cdot e^{-\rho-i\sigma}\, S^a_1$ OPE. The former calculation is relatively straightforward, while the latter requires some care; we explain some of the details in Appendix~\ref{app:symgen}. Note that this modified zero mode (\ref{fermsymmod}), together with the other zero modes, still obey the correct global $\mathfrak{psu}(1,1|2)$ commutation relations.


\section{Free field realisation}\label{FF}

It was shown in \cite{Eberhardt:2018ouy} that the $k=1$ version of the ${\rm AdS}_3 \times {\rm S}^3 \times \mathbb{T}^4$ hybrid string is dual to the symmetric orbifold of $\mathbb{T}^4$. Actually, the physical state condition for the hybrid string theory was not directly worked out in \cite{Eberhardt:2018ouy}. Instead, the equivalence of the hybrid description to the NS-R description for generic values of $k$ \cite{Troost:2011fd,Gaberdiel:2011vf,Seb} was used to determine what the physical state condition must amount to for the hybrid theory, and this was then extrapolated to the $k=1$ case. However, given the significance of this worldsheet theory, it would be very important to work out the cohomology of the $k=1$ theory directly. 

The level $k=1$ theory is most easily described in terms of a free field realisation of symplectic bosons and fermions \cite{Eberhardt:2018ouy} with (anti)-commutators 
\be\label{freefields1}
\{\psi^{\alpha}_r,\chi^\beta_s\} = \epsilon^{\alpha\beta}\, \delta_{r,-s} \ , \qquad 
[\xi^\alpha_r,\eta^\beta_s] = \epsilon^{\alpha\beta}\, \delta_{r,-s}  \ , 
\ee
where $\alpha,\beta\in\{\pm\}$, and $\epsilon^{\alpha\beta}$ is the anti-symmetric tensor with $\epsilon^{+-} = - \epsilon^{-+}=+1$, see Appendix~\ref{ffpsu1} for more details. As a first step towards determining its cohomology from first principles, we therefore need to express the relevant BRST operators in this free field language. This would appear to be a relatively straightforward exercise, but there is actually one important subtlety that we need to deal with (and that will occupy us for most of this paper): the free fields do not actually realise directly $\mathfrak{psu}(1,1|2)_1$, but rather $\mathfrak{u}(1,1|2)_1$, see Appendix~\ref{ulevk} for our conventions, and in order to obtain $\mathfrak{psu}(1,1|2)_1$ one needs to gauge the $\mathfrak{u}(1)$ field $Z$, see \cite{Eberhardt:2018ouy,DGGK}. For example, the central charge of the free fields equals $c=0$ (as befits a $\mathfrak{u}(1,1|2)_k$ theory), and the gauging by $Z$ reduces the central charge to $c=-2$, the central charge of the $\mathfrak{psu}(1,1|2)_k$ theory.\footnote{Gauging here means that we restrict to the states that are annihilated by the modes $Z_n$ with $n\geq 0$; since $[Z_m,Z_n]=0$, the $Z_{-n}$ descendants with $n>0$ are then null, and thus this gauging effectively removes two bosonic degrees of freedom and hence reduces the central charge by two.} As we saw above the central charge actually plays a critical role in the above OPEs, since for example the vanishing of the terms in (\ref{Q0Q0}) hinge on $T_{\rm int}$ having  central charge $c=-2$, see the discussion below eq.~(\ref{2.11}) and below eq.~(\ref{Rdef}). 

To be more specific, let us denote the stress energy tensor of the free field theory by 
\begin{equation}
\begin{aligned}
T_{\rm free}=& : \left(\dfrac{1}{2}\partial \xi^{-} \eta^{+}-\dfrac{1}{2}\xi^{-} \partial \eta^{+}-\dfrac{1}{2}\partial \xi^{+}\eta^{-}+\dfrac{1}{2}\xi^{+} \partial \eta^{-}\right) : \\
& + : \left(-\dfrac{1}{2}\partial \chi^{+}\psi^{-}-\dfrac{1}{2}\partial \psi^{-} \chi^{+}+\dfrac{1}{2}\partial \chi^{-} \psi^{+}+\dfrac{1}{2}\partial \psi^{+} \chi^{-}\right): \ ,
\label{tfree}
\end{aligned}
\end{equation}
where $:\cdots :$ denotes the usual normal ordering of free fields. For example, in the NS sector where all fields are half-integer moded, it is simply defined by 
\begin{equation}
\begin{aligned}
: a_r b_s : \, = \left\{ \begin{array}{ll}    ~~a_r b_s & \hbox{if $r \leq -\frac{1}{2}$} \\
                                                          \pm b_s a_r &  \hbox{if $r \geq \frac{1}{2}$} \ ,
                   \end{array} \right.
\label{CONCANO}
\end{aligned}
\end{equation}
where $a_r$, $b_s$ are any of the modes of the symplectic bosons or fermions, and the sign in the second line is `$-$' if both $a$ and $b$ are fermionic, and `$+$' otherwise. In the R sector where all fields are integer moded, we supplement this definition by setting
\be\label{Rnormal}
: a_0 b_0 : = \tfrac{1}{2} \bigl( a_0 b_0 \pm b_0 a_0 \bigr) \ . 
\ee
Then this normal ordering agrees with the so-called conformal normal ordering $\altcolon \cdot \altcolon$ that we shall work with in Appendix~\ref{NO}, see also \cite{Pol}. The reason why we consider the conformal normal ordering there is that this allows one to define the normal ordering of more than two fields in a straightforward manner, and this will be useful for our explicit computations. 

 Actually, it is not difficult to see that the free field stress energy tensor (\ref{tfree}) agrees with that of $\mathfrak{u}(1,1|2)_1$ which we may take to be given in terms of the Sugawara construction\footnote{The usual Sugawara construction is not applicable in this case, since the Killing form of $\mathfrak{u}(1,1|2)$ is zero \cite{BVW}. However, one can use the Halpern-Kiritsis construction \cite{Halpern:1989ss} and fix the remaining ambiguity by requiring that all the currents of $\mathfrak{u}(1,1|2)_1$ have spin one.}
\begin{equation}
\begin{aligned}
T_{\mathfrak{u}(1,1|2)}=&-:J^{3}J^{3}:+:\tfrac{1}{2}(J^{+}J^{-}+J^{-}J^{+}):+:K^{3}K^{3}: + :\tfrac{1}{2}(K^{+}K^{-}+K^{+}K^{-}):\\
& -\tfrac{1}{4}:\Bigl(\left( S^{++}_{1}S^{--}_{2}+S^{--}_{1}S^{++}_{2}-S^{+-}_{1}S^{-+}_{2}-S^{-+}_{1}S^{+-}_{2}\right) - (1 \leftrightarrow 2) \Bigr): \\
& -:ZY:+2:Z^{2}: \ ,
\label{tu}
\end{aligned}
\end{equation}
where all expressions are normal ordered as in eq.~(\ref{normalord}), and we use the conventions of Appendix~\ref{ulevk}.  This is to say, if we express the $\mathfrak{u}(1,1|2)_1$ generators in terms of the free fields as in Appendix~\ref{ffpsu1}, we find 
\be
T_{\rm free} = T_{\mathfrak{u}(1,1|2)} \ . 
\ee
One way to see this is to note that both $T_{\mathfrak{u}(1,1|2)}$ and $T_{\rm free}$ have the same OPEs with all the currents of $\mathfrak{u}(1,1|2)_1$. Thus their difference commutes with all currents and hence must be null. 
We have also checked this statement explicitly, following the analysis in \cite{Goddard, Pol}.

\subsection{The ${\cal N}=2$ generators in the free field realisation}

Next we observe that the first two lines in (\ref{tu}) formally look like the stress energy tensor of $\mathfrak{psu}(1,1|2)_1$, see eq.~(\ref{A.16}). However, this is a bit deceptive since the OPEs of the $S$ currents in $T_{\mathfrak{u}(1,1|2)}$, see eq.~(\ref{eq:ss-z}), differ from those in $\mathfrak{psu}(1,1|2)_1$ by the $Z$-terms. Thus it is not directly possible to extract the $\mathfrak{psu}(1,1|2)_1$ stress energy tensor from these expressions. One may nevertheless be tempted to define the `fake $\mathfrak{psu}(1,1|2)_1$ stress tensor' by 
\begin{equation}
\begin{aligned}
T_{\mathfrak{psu}}^{({\rm f})}=T_{\rm free} + \left(:ZY:-2:Z^{2}: \right) \ ,
\label{psunfree}
\end{aligned}
\end{equation}
but it actually does not satisfy the OPEs of a stress energy tensor, 
\be
T_{\mathfrak{psu}}^{({\rm f})} (z) \, T_{\mathfrak{psu}}^{({\rm f})} (w) = \frac{-1}{ (z-w)^4} + \frac{2 \, T_{\mathfrak{psu}}^{({\rm f})}(w)+4Z^2(w)}{(z-w)^2} + \frac{\partial T_{\mathfrak{psu}}^{({\rm f})}(w)+2\partial Z^2(w)}{(z-w)} + {\cal O}\bigl( (z-w)^0 \bigr) \ , 
\ee
and hence cannot be used as $T_{\rm int}$ in the eqs.~(\ref{Tdef}) and (\ref{GPdef}). Instead, we will set 
\be\label{Tint}
T_{\rm int}=T_{\rm free}+ZY \ . 
\ee
Note that this is effectively the coset stress energy tensor since 
\be
T_{UV} = - ZY = V^2 - U^2 
\ee
is the stress energy tensor of the $(U,V)$ system. In particular, $T_{\rm int}$ therefore commutes with the modes of $U$ and $V$, and hence defines a stress energy tensor with $c=-2$, as is appropriate for 
$T_{\rm int}$. In particular, this then guarantees that the OPE $Q_{0}Q_{0}$ is regular, because $T_{\rm int}$ has the right central charge. 

The other term that requires some care is the $R$-term in $G^+$, see eq.~(\ref{Rdef}). The details of this calculation are explained in Appendix~\ref{Rterm}, and it leads to 
\begin{equation}
R = - \psi^{+}\psi^{-}(\eta^{+}\partial \eta^{-}-\eta^{-} \partial \eta^{+}) \ , \qquad 
Q_{1}=-e^{-\rho}\psi^{+}\psi^{-}(\eta^{+}\partial \eta^{-}-\eta^{-} \partial \eta^{+})\ .
\label{Q1free}
\end{equation}
This agrees with eq.~(3.6) of \cite{DGGK}, up to a factor of 2. Relative to  \cite{DGGK}  we have also reversed the roles of the two sets of fermions and symplectic bosons. 

With these definitions of $T_{\rm int}$, see eq.~(\ref{Tint}), and $Q_1$, see eq.~(\ref{Q1free}), we can then evaluate the OPEs of the generators defined in eq.~(\ref{N2gencur}). Unfortunately, they do not satisfy an ${\cal N}=2$ algebra any longer, since the two OPEs
\begin{align}
T(z)G^{+}(w) & \sim \frac{G^{+}(w)}{(z-w)^{2}}+\frac{\partial_{w} G^{+}}{z-w} + 
\frac{2e^{-\rho}:ZR:(w)}{z-w} \ , \label{mis1}\\
G^{+}(z)G^{+}(w) & \sim -\frac{4e^{-\rho+i\sigma}:ZR:(w)}{z-w}  \label{mis2}
\end{align}
have correction terms proportional to $:ZR:$. This is not completely unexpected since the free fields only generate the $\mathfrak{u}(1,1|2)_1$ algebra, and this differs from $\mathfrak{psu}(1,1|2)_1$ by terms proportional to $Z$, see eq.~(\ref{eq:ss-z}). We mention in passing that these correction terms are a consequence of the fact that $R$ defined in (\ref{Q1free}) is not a primary field of weight $3$ w.r.t.\ 
$T_{\rm int}=T_{\rm free}+ZY$ since it has a non-trivial OPE with $Y$; more specifically we have  
\be
T_{\rm int}(z) \, R(w) \sim  \frac{3R(w)}{(z-w)^2}+\frac{\partial_wR}{z-w}+\frac{2:ZR:(w)}{z-w} \ . 
\ee

 \subsection{The global symmetry generators} \label{sec:global-free}
 
 A similar problem also arises for the global symmetry generators  $(\tilde{S}^a_2)_0$, see eq.~(\ref{fermsymmod}) in Section~\ref{sec:global}. (The other set of fermionic generators, $({S}^a_1)_0$, continue to commute with the free field versions of  eq.~(\ref{N2gencur}).) Expressing $(\tilde{S}^a_2)_0$ in terms of the free fields we find 
\be
q^{\alpha\beta} = \oint \left( \xi^{\alpha}\chi^{\beta}+e^{-\rho-i\sigma}\eta^{\alpha}\psi^{\beta} \right)\ ,
\label{5.2}
\ee
where we now label them in terms of $(\alpha\beta) \cong a$.
The anti-commutator of $q^{++}$ with $G^+$ --- here we use the free field version of eq.~(\ref{GPdef}) with $T_{\rm int}$ and $Q_1$ defined in eqs.~(\ref{Tint}) and (\ref{Q1free}), respectively --- then equals 
\begin{equation}
\begin{aligned}
& \{q^{++},G^{+}\}   \\
& \quad = \, e^{-\rho}\left[-\partial(\rho+i\sigma):ZS^{++}_{1}:-\altcolon\eta^{+}\psi^{+}T_{\rm free}\altcolon-\tfrac{1}{4}\altcolon\eta^{+}\psi^{+}(\eta^{+}\xi^{-}-\eta^{-}\xi^{+})^{2}\altcolon\right.\\
&\qquad +\tfrac{1}{4}\altcolon\bigl( \partial\psi^{+}\eta^{+}(-2\eta^{+}\xi^{-}+2\eta^{-}\xi^{+}-\psi^{+}\chi^{-})+\psi^{+}\partial\eta^{+}(-4\xi^{+}\eta^{-}+\eta^{+}\xi^{-}+2\chi^{+}\psi^{-})\bigr)\altcolon\\
&\qquad +\tfrac{1}{4}\altcolon\bigl(\eta^{+}\psi^{+}(-\partial\xi^{-}\eta^{+}+3\partial\eta^{-}\xi^{+}+\partial\xi^{+}\eta^{-}-\partial\psi^{-}\chi^{+}-5\partial\chi^{+}\psi^{-}) \bigr)\altcolon\\
&\qquad \left.+\tfrac{1}{8}\altcolon(\partial^{2}\eta^{+}\psi^{+}-\partial^{2}\psi^{+}\eta^{+})\altcolon\right] \ , 
\label{5.3}
\end{aligned}
\end{equation}
where for the terms that involve more than two fields we have used the conformal normal ordering prescription of Appendix~\ref{NO}. This expression is again proportional to $Z$ since we may write it as 
\begin{equation}
\begin{aligned}
\{q^{++},G^{+}\}  = e^{-\rho}\Bigl(&-\partial(\rho+i\sigma):ZS^{++}_{1}:+:Z(J^{3}+K^{3})S^{++}_{1}:+:ZJ^{+}S^{-+}_{1}:\\
& +:ZK^{+}S^{+-}_{1}: -2:S^{++}_{1}Z^{2}:+:\partial ZS^{++}_{1}:-2:Z\partial S^{++}_{1}:\Bigr)\ .
\label{deldel}
\end{aligned}
\end{equation}
Thus if we quotient out by $Z$ suitably, the generator \eqref{5.2} will commute with $G^+$. One way to achieve this will be explained in the next section. 


\section{An $\mathcal{N}=4$ algebra for the free-field representation}\label{sec:4}

As we have seen in the previous section, it does not seem possible to construct an actual ${\cal N}=2$ algebra in terms of the free fields of  the level $k=1$ theory. The reason for this difficulty is that the free fields only realise $\mathfrak{u}(1,1|2)_1$, and that in order to reduce this to $\mathfrak{psu}(1,1|2)_1$ one needs to gauge by the $Z$ field. As a consequence, this $Z$ field `contaminates' the ${\cal N}=2$ relations.

In this section we want to explain how we can recover an honest ${\cal N}=2$ algebra if we add additional ghost fields that incorporate the gauging of the $Z$ field. We shall also argue that the corresponding BRST cohomology imposes then both the $Z$ gauging condition, as well as the physical state condition of the hybrid string theory. Finally we shall show that the global $\mathfrak{psu}(1,1|2)$ generators commute with the ${\cal N}=2$ algebra (as well as the extended ${\cal N}=4$ algebra we are also about to discuss, see eqs.~(\ref{eq:su2-n=4}) and (\ref{eq:n=4-added}) below) on the corresponding physical states. This extended set-up, including the additional ghosts that we are about to introduce, is therefore a good starting point for further discussions of this worldsheet theory.

\subsection{An actual $\mathcal{N}=2$ algebra} \label{sec:n=4-modified}

As we have explained above, the free field theory of the symplectic bosons and fermions does not quite lead to the $\mathfrak{psu}(1,1|2)_1$ worldsheet theory we need, but rather to an $\mathfrak{u}(1,1|2)_1$ theory. In order to reduce this to $\mathfrak{psu}(1,1|2)_1$  we need to gauge the $Z$ field, see e.g.\ eq.~(\ref{eq:ss-z}). This gauging can be performed by introducing additional ghost fields (and $\mathfrak{u}(1)$ fields), and defining a suitable BRST operator as in \cite{Karabali:1989dk}. Our setting is slightly different, however, since we want to describe the cohomology in terms of a topological ${\cal N}=4$ string, and thus the BRST operators should come from the $G^+$ supercurrents. As a consequence, we cannot literally follow the construction of \cite{Karabali:1989dk}: if we were to modify $T_{\rm int}$ by replacing the $ZY$ term in (\ref{Tint}) by $Z'Y'$, where $Z'$ and $Y'$ are the additional $\mathfrak{u}(1)$ fields, and then simply add $c (Z+Z')$ to $G^+$, where $c$ is one of the additional ghosts,  then $R$ would be primary of weight $3$ w.r.t.\ the new stress energy tensor, but the $G^+ G^+$ OPE would still not be zero, but rather equal
\be
G^+(z) \, G^+(w) \sim \frac{\partial_w(e^{i\sigma}c(Z+Z'))}{z-w} \ .
\ee
We shall therefore follow a slightly different route in that we shall not directly use the construction of 
\cite{Karabali:1989dk}, but rather be inspired by their general setup. More specifically, we propose to add the following ghost fields to our set-up: 
%
\begin{itemize}
\item an anti-commuting ghost $(b,c)$ with weights $(1,0)$ and OPE $b(z) c(w) \sim \frac{1}{z-w}$ \ ,
\item an anti-commuting ghost $(b^{\prime},c^{\prime})$ with weights $(1,0)$ and OPE $b'(z) c'(w) \sim \frac{1}{z-w}$ \ ,
\item a commuting ghost $(\beta^{\prime},\gamma^{\prime})$ with weights $(1,0)$ and OPE $\beta^\prime(z) \gamma^\prime(w) \sim \frac{1}{z-w}$\ .
\end{itemize}
We note that these ghosts have central charges $-2$, $-2$ and $2$, respectively, and therefore their sum is indeed $c=-2$, as desired. Here the $(b,c)$ and $(b^{\prime},c^{\prime})$ play the role of the ghost fields of \cite{Karabali:1989dk}, while $(\beta^{\prime},\gamma^{\prime})$ mimics the additional $\mathfrak{u}(1)$ generators $Z'$ and $Y'$ -- in the language of \cite{Karabali:1989dk}, these are associated to the $H\cong {\rm U}(1)\times {\rm U}(1)$ subgroup that is being gauged. We then claim that the following generators satisfy a  twisted ${\cal N}=2$ superconformal algebra:
%
\begin{subequations} \label{eq:exact-psu-1-generators}
	\begin{align}
	\begin{split} \label{eq:exact-t}
	T & = \hat{T}_{\rm int} - \frac{1}{2} \partial \rho \partial \rho - \frac{1}{2} \partial \sigma \partial \sigma 
+ \frac{3}{2} \partial^2( \rho+i\sigma) + T_{C} 
	\end{split}\\
\begin{split}\label{G+exact}
	G^+ & = e^{-\rho} R
+ e^{i\sigma} \Bigl( \hat{T}_{\rm int} - \frac{1}{2} \partial (\rho + i \sigma) \partial (\rho + i \sigma) + \frac{1}{2} \partial^2 (\rho + i \sigma) \Bigr)
+ G^+_C
\end{split}\\
\begin{split}
	G^- &= 2 e^{-i\sigma} + G^-_{C}
\end{split}\\
\begin{split}
J & = \partial (\rho + i \sigma) + J_C \label{eq:j} \ .
\end{split}
	\end{align}
\end{subequations}
Here, the new internal stress energy tensor $\hat{T}_{\text{int}}$ equals
\begin{equation} \label{eq:tint-modified}
\hat{T}_{\text{int}}=T_{\mathfrak{u}(1,1|2)} + T_{bc} + T_{b^{\prime}c^{\prime}}+T_{\beta^{\prime}\gamma^{\prime}} + \partial (\gamma^{\prime} Z) \ ,
\end{equation}
where $T_{\mathfrak{u}(1,1|2)}=T_{\rm free}$, see eqs.~(\ref{tfree}) and (\ref{tu}), while 
$T_{bc}$, $T_{b^{\prime}c^{\prime}}$ and $T_{\beta^{\prime}\gamma^{\prime}}$ are the respective ghost stress-tensors, i.e.\
\begin{equation} \label{eq:ghosts-t}
T_{bc}=((\partial c)b) \ , \qquad T_{b^{\prime}c^{\prime}}=((\partial c^{\prime})b^{\prime}) \ , \qquad T_{\beta^{\prime}\gamma^{\prime}}=((\partial \gamma^{\prime})\beta^{\prime}) \ . 
\end{equation}
Furthermore, the $R$ field in $G^+$ is defined as in eq.~(\ref{Q1free}). Note that the $Q_2$ term in $G^+$ of Section~\ref{curve}, see eqs.~(\ref{GPdef}) and (\ref{S4def}), actually vanishes in the free field realisation, as was already noted in \cite{DGGK}. 

It is relatively straightforward to show that $\hat{T}_{\text{int}}$ of eq.~(\ref{eq:tint-modified}) defines a stress-tensor with central charge $c=-2$. Moreover, the OPE of $R$ with $Z$ is trivial, and therefore $R$ is a primary of weight $3$ with respect to $\hat{T}_{\text{int}}$ ---  note that $\hat{T}_{\text{int}}$ does not involve the $ZY$ term of (\ref{Tint}) any longer. It then follows from the analysis in Section~\ref{sec:flat} that the fields in eqs.~(\ref{eq:exact-psu-1-generators}) define a twisted $\mathcal{N}=2$ algebra; since the $Q_2$ term is absent in the free field realisation and the $Q_1 Q_1$ OPE is regular, all the terms in (\ref{Q2Q0}) are individually trivial.

We mention in passing that there are also other definitions that would have led to a consistent ${\cal N}=2$ algebra; for example, we could have corrected the $R$ term (instead of adding the $Z$-dependent correction term to $T_{\rm int}$). We could have also added $\gamma^{\prime}Z$ instead  to $J$ in eq.~(\ref{eq:j}) and modified $G^+$ by adding $-\partial(e^{i\sigma} \gamma^{\prime}Z)$.
\medskip


As in \cite{BVW} we can now enhance this ${\cal N}=2$ algebra to an ${\cal N}=4$ algebra by considering the currents 
\begin{equation} \label{eq:su2-n=4}
J^{\pm\pm}=e^{\pm(\rho+i\sigma+iH)}
\end{equation}
that extend the $\mathfrak{u}(1)$ symmetry generated by $J$, to an $\mathfrak{su}(2)$ symmetry. The commutator of the corresponding zero modes with the supercurrents (see eqs.~(\ref{N4twis})), then lead to the additional supercurrents\footnote{For the compact part, the supercurrents $\tilde{G}^+$ and $\tilde{G}^-$ are defined using a boson $H$ with the OPE $H(z) H(w) \sim -2 \ln(z-w)$. The $\mathfrak{su}(2)_1$ currents are then $J^{\pm\pm}_C= e^{\pm iH}$ and $J_C=i\partial H$, see eqs.~(\ref{eq:j-j-su2}), (\ref{eq:j-jpm-su2}) and (\ref{eq:jp-jm-su2}).} 
\begin{subequations} \label{eq:n=4-added}
\begin{align} \label{eq:gtilde+}
	\tilde{G}^+ & = 2 e^{\rho+i H} + e^{\rho+i\sigma} \tilde{G}^+_C \ , \\ 
	\tilde{G}^{-} & = e^{-2\rho-i \sigma - i H} R - e^{-\rho - i H}\Bigl( \hat{T}_{\rm int} - \frac{1}{2} \partial (\rho + i \sigma) \partial (\rho + i \sigma) + \frac{1}{2} \partial^2 (\rho + i \sigma) \Bigr) + e^{-\rho-i\sigma} \tilde{G}^{-}_C \ .
\end{align}
\end{subequations}
These fields together then generate a twisted $\mathcal{N}=4$ algebra, see Appendix~\ref{N2} for our conventions. (Note that the fields $J^{\pm\pm}$ commute with $\gamma^{\prime} Z$, and thus their conformal dimension is unmodified by the $\gamma^{\prime} Z$ correction term in $T_{\rm int}$.)

As in \cite{BVW} we can then use this ${\cal N}=4$ algebra to define the BRST cohomology that characterises the physical states. This will be discussed in more detail in the following section. 

%

\subsection{Cohomology and $Z_n=0$ condition}\label{sec:modifiedcohomology}

Recall from \cite{BVW} that the physical states are characterised by the cohomology conditions 
%
%
\begin{equation} \label{eq:cohomology-conditions}
G^+_0 \psi = \tilde{G}^+_0 \psi = (J_0-1) \psi = T_0 \psi = 0 \ , \quad\quad \psi \sim \psi + G^+_0 \tilde{G}^+_0 \psi^{\prime} \ .
\end{equation}
We now propose that in our free field set-up we should do the same. This is to say, we define the ${\cal N}=4$ generators in terms of the free fields and the additional ghost fields as discussed in the previous section. We then claim that the physical states of our level $k=1$ free field description are characterised by the  cohomology of (\ref{eq:cohomology-conditions}). 

We now want to argue that this will describe the correct physical spectrum. In particular, it should imply that the physical states are annihilated by $Z_n$ with $n\geq 0$, i.e.\ that they are part of  the $\mathfrak{psu}(1,1|2)_1$ theory. As we argue below, this should then be sufficient to guarantee that the resulting cohomology will coincide with that of \cite{BVW} as formulated directly in terms of the $\mathfrak{psu}(1,1|2)_k$ WZW model. 

Unfortunately, a complete BRST analysis of (\ref{eq:cohomology-conditions}) is quite difficult (although, in this free field set-up, this is now maybe within reach); note that also the cohomology of \cite{BVW} has only been very partially studied \cite{Troost:2011fd,Gaberdiel:2011vf,Seb}, and that a complete description of the physical spectrum in this language is still missing. In order to make progress we shall assume that we can choose a representative in each cohomology class that satisfies\footnote{Note that spectral flow does not act on the ghost fields, and thus these conditions apply equally to the spectrally flowed sectors. The same also applies to the conditions in eq.~(\ref{stronger}) below; in particular, $Z_n$ is not affected by spectral flow, and the spectrally flowed $T_0 \psi=0$ or $\textbf{T}_0 \psi=0$ condition is precisely what is required to get the correct physical spectrum following \cite{Gaberdiel:2018rqv} and \cite{Eberhardt:2018ouy}.}
\begin{equation} \label{eq:ghosts-conditions}
b_n \psi = b'_n \psi = \beta'_n \psi =0 \quad \hbox{for $n \geq 0$} \ , \quad\quad\quad 
c_n \psi = c'_n \psi = \gamma'_n \psi =0 \quad \hbox{for $n \geq 1$} \ .
\end{equation}
This seems reasonable since these conditions only refer to the additional ghosts we have added to our description. We now want to show that provided this is the case, i.e.\ provided that $\psi$ is in the BRST cohomology of eq.~(\ref{eq:cohomology-conditions}) and satisfies eq.~(\ref{eq:ghosts-conditions}), then $\psi$ also satisfies 
\be\label{stronger}
\textbf{G}^+_0 \psi =  \textbf{T}_0\, \psi =  Z_n \psi = 0 \qquad n \geq 0 \ , 
\ee
where 
%
%
\begin{subequations} \label{eq:non-exact-psu-1-generators}
\begin{align}
\begin{split} \label{eq:non-exact-t}
\textbf{T} & = T_{{\rm free}} - \frac{1}{2} \partial \rho \partial \rho - \frac{1}{2} \partial \sigma \partial \sigma + \frac{3}{2} \partial^2( \rho+i\sigma) + T_{C} 
\end{split}\\
\begin{split}
\textbf{G}^+ & = e^{-\rho} R
+ e^{i\sigma} \Bigl( T_{{\rm free}} - \frac{1}{2} \partial (\rho + i \sigma) \partial (\rho + i \sigma) + \frac{1}{2} \partial^2 (\rho + i \sigma) \Bigr)
+ G^+_C 
\end{split}
\end{align}
\end{subequations}
are the `old' ${\cal N}=2$ generators without the additional ghosts and the $(\gamma' Z)$ correction term in $\hat{T}_{\rm int}$. In particular, this should mean that they are part of the original cohomology of \cite{BVW}.

In order to show eq.~(\ref{stronger}), we note that eq.~(\ref{eq:ghosts-conditions}) in particular means that 
\be\label{eq:tfg=n}
(T_{bc})_n \psi = (T_{b^{\prime}c^{\prime}})_n \psi = (T_{\beta^{\prime}\gamma^{\prime}})_n \psi = 0 \ , \qquad n\geq -1 \ . 
\ee
Thus the BRST condition $T_0\psi=0$, see eq.~(\ref{eq:cohomology-conditions}), implies that 
\begin{equation} \label{eq:z-from-t}
T_0 \, \psi = \textbf{T}_0\,  \psi - (\gamma^{\prime} Z)_0 \psi = 0 \ ,
\end{equation}
where $\textbf{T}$ is given in eq.~(\ref{eq:non-exact-t}), and we have used eq.~(\ref{eq:tfg=n}) with $n=0$ for each of the additional ghost terms. We can write out the correction term in eq.~(\ref{eq:z-from-t}) as 
\begin{equation}
(\gamma^{\prime} Z)_0 \psi = \sum_{j=0}^{\infty} \gamma^{\prime}_{-j} Z_j \psi \ .
\end{equation}
Applying $\beta^{\prime}_{n}$ with $n\geq 0$ to eq.~(\ref{eq:z-from-t}), and using that $\beta^{\prime}_{n}$ commutes with ${\bf T}_0$ as well as eq.~(\ref{eq:ghosts-conditions}), we therefore deduce that 
\begin{equation}\label{Zncond}
Z_n \psi = 0 \qquad n\geq 0 \ ,
\end{equation}
thereby giving the last identity in eq.~(\ref{stronger}). In fact, also the second identity in eq.~(\ref{stronger}) is now manifest since the correction term in eq.~(\ref{eq:z-from-t}) vanishes, and hence 
\begin{equation}
	\textbf{T}_0 \, \psi = 0 \ .
\end{equation}
It remains to show that $\psi$ is also annihilated by $\textbf{G}^+_0$. Using that the full $G^+_0$ annihilates $\psi$, see eq.~(\ref{eq:cohomology-conditions}), we have 
\begin{equation} \label{eq:difference-g}
	G^+_0 \psi = \textbf{G}^+_0 \psi + \sum_{n=2}^{\infty} \textbf{c}_n T^{\text{extra}}_{-n} \psi = 0 \ ,
\end{equation}
where $\textbf{c}=e^{i\sigma}$ is the $c$-ghost of the original hybrid formulation\footnote{This is not to be confused with the additional ghosts we have introduced above.} and
\begin{equation} \label{eq:textra}
	T^{\text{extra}}= T_{bc} + T_{b^{\prime}c^{\prime}}+T_{\beta^{\prime}\gamma^{\prime}} + \partial (\gamma^{\prime} Z) \ .
\end{equation}
Here we have used that eqs.~(\ref{eq:tfg=n}) and (\ref{Zncond}) imply
\begin{equation} \label{eq:textra=0}
	T^{\text{extra}}_n \psi = 0 \quad \quad n \geq -1 \ ,
\end{equation}
and hence that only the negative modes less than $-1$ of $T^{\text{extra}}$ enter in eq.~(\ref{eq:difference-g}). 
Furthermore, using the gauge equivalence in eq.~(\ref{eq:cohomology-conditions}) we should be able to find a representative $\psi$ for which 
\begin{equation} \label{eq:b-c-condition}
\textbf{c}_n \psi = 0 \qquad n \geq 1 \ .
\end{equation}
Thus the remaining terms in eq.~(\ref{eq:difference-g}) vanish, and we conclude that we can always find a representative for which $\textbf{G}^+_0 \psi = 0$. This completes our argument. 

%
%

\subsection{Global symmetry}

Finally, we need to address the question of whether the global symmetries in $\mathfrak{psu}(1,1|2)$ commmute with the ${\cal N}=2$ algebra, see the discussion in Sections~\ref{sec:global} and \ref{sec:global-free}. As was explained in Section~\ref{sec:global}, one would naively expect the zero modes of $\mathfrak{psu}(1,1|2)_k$ to commute with the ${\cal N}=2$ generators. While this is true on the nose for all the bosonic and half of the fermionic generators, we had to modify half of the fermionic generators so as to make them (anti)-commute, see eq.~(\ref{fermsymmod}).  Furthermore, the construction broke down again in the free field realisation since there were $Z$ dependent correction terms, see eq.~(\ref{deldel}). Here we will explain that while these $\mathfrak{psu}(1,1|2)$  generators do not (anti)-commute with the $\mathcal{N}=4$ generators in the free-field realisation (i.e.\ the generators in eqs.~(\ref{eq:exact-psu-1-generators}), eq.~(\ref{eq:su2-n=4}) and eqs.~(\ref{eq:n=4-added})), these (anti)-commutators vanish on physical states, and hence the physical states sit in representations of $\mathfrak{psu}(1,1|2)$.

Let us start by noting that our ${\cal N}=2$ generators only differ from those in Section~\ref{sec:global-free} by the fact that we have modified the stress energy tensor, i.e.\ replaced 
\be
T_{\rm int} \rightarrow \hat{T}_{\rm int} = T_{\rm int} - ZY + T_{bc} +T_{b'c'} + T_{\beta'\gamma'} + \partial (\gamma' Z) \ . 
\ee
Since all the generators of $\mathfrak{psu}(1,1|2)_1$ commute with $Z_n$ (and do not involve any of the additional ghost fields), the only interesting correction term is the $-ZY$ term. It has an effect since the fermionic generators do not commute with $Y_n$,
\begin{equation}
	[Y_n,(S^{\alpha\beta}_{i})_m]=(-1)^{i+1} (S^{\alpha\beta}_{i})_{n+m} \ ,
\end{equation}
see Appendix~\ref{ffpsu1}. However, since in the correction term $Y$ appears together with $Z$, the $ZY$ term behaves as\footnote{Note that in $G^+$, see eq.~(\ref{G+exact}), the relevant term involves also the ghost factor $e^{i\sigma}$, but this does not play a role here since it commutes with the correction terms we have added. However, it leads to a simple pole with the ghost factor $e^{-\rho-i\sigma}$ in the second term in eq.~(\ref{5.2}), but this pole is then proportional to $YZ$, and hence, in particular, proportional to $Z$.}
\be
[ (ZY)_n,(S^{\alpha\beta}_{i})_m]= (-1)^{i+1} (ZS^{\alpha\beta}_{i})_{n+m} \ .
\ee
As a consequence, the right-hand-side of (\ref{deldel}) will still only involve terms proportional to $Z$, and they are removed by the gauge condition $Z_n \psi=0$ for $n\geq 0$, see eq.~(\ref{stronger}), resp.\ are null because $Z_m^\dagger = - Z_{-m}$. 

Finally, we note that the global symmetry generators also commute with the $J^{\pm\pm}$ generators of eq.~(\ref{eq:su2-n=4}) since the latter do not involve any $\mathfrak{psu}(1,1|2)_1$ generators, and the ghost factor $e^{-\rho-i\sigma}$ in eq.~(\ref{5.2}) has a regular OPE with $J^{\pm\pm}$. Thus, the global symmetry generators also commute on physical states with the $\mathcal{N}=4$ algebra generated by these fields, and this is sufficient to deduce that the physical states must sit in representations of the global symmetry algebra $\mathfrak{psu}(1,1|2)$, as expected.

\section{Conclusion}\label{sec:concl}

In this paper we have undertaken first steps towards writing the BVW hybrid worldsheet theory \cite{BVW} for the pure $k=1$ NS-NS background in terms of free fields, using the free field realisation of the $\mathfrak{psu}(1,1|2)_1$ affine algebra. As we stressed throughout, the main subtlety with this construction is related to the fact that the free fields only realise $\mathfrak{u}(1,1|2)_1$, and that one needs to gauge the $\mathfrak{u}(1)$ $Z$-field  in order to obtain actually $\mathfrak{psu}(1,1|2)_1$. In particular, this is responsible for contaminating the ${\cal N}=2$ algebra relations that underpin the BVW construction in the (naive) free field realisation.

As we explained in Section~\ref{sec:4} it is possible to overcome these difficulties by introducing an additional ghost system that implements the $Z$ gauging, following at least in spirit the construction of \cite{Karabali:1989dk}. In particular, this allowed us to construct an ${\cal N}=2$ algebra in our free field setting, and it seems plausible that the cohomology of the associated ${\cal N}=4$ topological string will lead to the correct physical spectrum, see Section~\ref{sec:modifiedcohomology}. 

There are a number of natural future directions. First of all, it would be very important to work out the cohomology of our ${\cal N}=4$ system of Section~\ref{sec:4} from first principles; given that this is now a free field construction, this should be feasible, and it would be very satisfying if this could be shown to reproduce what is expected from the arguments in \cite{Eberhardt:2018ouy}. In turn, this should also allow us to work out the correlators of \cite{DGGK} more fully --- in particular, it should then be possible to determine the actual ghost contributions and hence check whether the undetermined coefficients in \cite{DGGK} reproduce what is expected from the dual symmetric orbifold, see e.g.\ \cite{Lunin:2001}. 

The other natural direction is to try and generalise these considerations to the worldsheet theory that has been proposed to be exactly dual to free SYM in 4D \cite{Gaberdiel:2021iil,Gaberdiel:2021jrv}. This worldsheet theory essentially consists of twice the symplectic bosons and fermions, but also needs to be quotiented out by the analogue of the $Z$-field --- this was denoted by ${\cal C}$ in \cite{Gaberdiel:2021iil,Gaberdiel:2021jrv}. One should therefore be able to construct an ${\cal N}=4$ superconformal algebra for this theory, following essentially the same arguments as in this paper, and it would be interesting to see whether this can be further extended to ${\cal N}=8$.

\acknowledgments

We thank Rajesh Gopakumar for useful discussions and collaboration at an early stage. We also thank Rajesh Gopakumar and Bob Knighton for comments on a draft version of this paper. The work of KN and VS is supported by a grant from the Swiss National Science Foundation. The activities of the group are more generally supported by the NCCR SwissMAP, which is also funded by the Swiss National Science Foundation.

\appendix
\section{Conventions}

In this appendix we collect various conventions.

\subsection{The $\mathcal{N}=2$ algebra and the topological twist}\label{N2}

Let us start by reviewing the topologically twisted $\mathcal{N}=2$ superconformal algebra with central charge $c$ and introduce our conventions. The fields of this algebra consist of the stress-tensor $T$ with modes $L_n$, a $\mathfrak{u}(1)$ current $J$ with modes $J_n$, and two supercurrents $G^{\pm}$ with modes $G^{\pm}_{n}$. Here we consider the usual topological twist where we add $\frac{1}{2} \partial J$ to the stress-tensor $T$ \cite{Witten:1988xj}. Then the resulting stress tensor has vanishing central charge, while the conformal dimensions of the primary fields are shifted according to their $\mathfrak{u}(1)$ charge, e.g.\ $G^+$ and $G^-$ will have weights $1$ and $2$, respectively. The resulting topologically twisted algebra has then the commutation relations
\begin{subequations}\label{N2twis}
\begin{eqnarray}
{}[L_m,L_n] & = & (m-n) L_{m+n} \\
\{ G^+_m, G^-_n \} & = & 2 L_{m+n} + 2 m J_{m+n} +\, \tfrac{c}{3} n (n+1) \, \delta_{m,-n} \\
{} [L_m,G^+_n] & = & - n \, G^+_{m+n} \\
{} [L_m,G^-_n] & = & (m-n) \, G^-_{m+n} \\ 
{} [ J_m,G^\pm_n] & = & \pm G^\pm _{m+n} \\ 
{} [L_m,J_n] & = & - n J_{m+n} - \tfrac{c}{6} m (m+1) \delta_{m,-n} \\
{} [J_m,J_n] & = & \tfrac{c}{3}\, m \, \delta_{m,-n} \label{eq:j-j-su2} \ . 
\end{eqnarray}
\end{subequations}
We shall also need the extension of this algebra to the topologically twisted ${\cal N}=4$ algebra; the latter has, in addition, the generators $J^{++}_m$, $J^{--}_m$ and $\tilde{G}^\pm_m$, with the non-trivial commutation relations being 
{\allowdisplaybreaks
\begin{subequations}\label{N4twis}
\begin{eqnarray}
{} [L_m,J^{++}_n] & = & -(m+n) \, J^{++}_{m+n} \\
{} [L_m,J^{--}_n] & = & (m-n) \, J^{--}_{m+n} \\
{} [L_m,\tilde{G}^{+}_n] & = & -n \, \tilde{G}^+_{m+n} \\
{} [L_m,\tilde{G}^{-}_n] & = & (m-n) \, \tilde{G}^-_{m+n} \\
{}[J_m,J^{\pm\pm}_n] & = & \pm \, 2\,  J^{\pm\pm}_{m+n} \label{eq:j-jpm-su2} \\ 
{} [J_m,\tilde{G}^{\pm}_n] & = & \pm \, \tilde{G}^{\pm}_{m+n} \\
{}[J^{++}_m,J^{--}_n] & = & J_{m+n} + \tfrac{c}{6} \, m \, \delta_{m,-n} \label{eq:jp-jm-su2} \\
{} [J^{\pm\pm}_m,G^{\mp}_n] & = & \pm \,\tilde{G}^{\pm}_{m+n} \\
{} [J^{\pm\pm}_m,\tilde{G}^{\mp}_n] & = & \mp \, G^{\pm}_{m+n} \\
{} [G^{\pm}_m,\tilde{G}^{\pm}_n] & = & \mp \, 2 \,(m-n) \, J^{\pm\pm}_{m+n} \\
\{ \tilde{G}^+_m, \tilde{G}^-_n \} & = & 2 \, L_{m+n} + 2 \, m \, J_{m+n} +\, \tfrac{c}{3} n (n+1) \, \delta_{m,-n} \ .
\end{eqnarray}
\end{subequations}}

\subsection{$\mathfrak{psu}(1,1|2)_{k}$ conventions}\label{psulevk}

The other important algebra that will play a significant role in our analysis is $\mathfrak{psu}(1,1|2)_k$. There are two natural conventions in which one may describe it, namely those of \cite{BVW}, and those of \cite{Eberhardt:2018ouy,DGGK}. We shall also spell out how these conventions are related to one another. 

In the conventions of \cite{BVW}, $\mathfrak{psu}(1,1|2)_k$ consists of the bosonic generators $K^{ab}_m = - K^{ba}_m$ and the fermionic generators $(S^a_\alpha)_m$, with commutation relations 
\begin{equation}
\begin{aligned}
{}[K^{ab}_{m},K^{cd}_{n}] & =mk\epsilon^{abcd}\delta_{m+n,0}+\delta^{ac}K^{bd}_{m+n}-\delta^{ad}K^{bc}_{m+n}-\delta^{bc}K^{ad}_{m+n}+\delta^{bd}K^{ac}_{m+n}\ , \\ 
\{(S^{a}_{\alpha})_{m},(S^{b}_{\beta})_{n}\} & =mk\delta^{ab}\delta_{m+n,0}\epsilon_{\alpha \beta}+\dfrac{1}{2}\epsilon_{\alpha \beta}\epsilon^{abcd}K^{cd}_{m+n}\ , \\ 
{}[K^{ab}_{m},(S^{c}_{\alpha})_{n}] & =\delta^{ac}(S^{b}_{\alpha})_{m+n}-\delta^{bc}(S^{a}_{\alpha})_{m+n} \  .
\label{3.19}
\end{aligned}
\end{equation}
Here the latin indices run over $a\in\{1,2,3,4\}$, while the greek indices take values in \linebreak $\alpha,\beta\in\{1,2\}$. Furthermore, the $\epsilon$ tensors are totally anti-symmetric with the convention that $\epsilon_{12}=1=\epsilon^{1234}$.

On the other hand, in the conventions of \cite{Eberhardt:2018ouy,DGGK}, the bosonic generators of $\mathfrak{psu}(1,1|2)_k$ are $J^a_m \in \mathfrak{sl}(2,\mathbb{R})_k$ and $K^a_m \in \mathfrak{su}(2)_k$, where now $a\in\{3,\pm\}$. The fermionic generators, on the other hand, are denoted by $S^{\alpha\beta\gamma}_m$, where now $\alpha,\beta,\gamma\in\{\pm\}$. In these conventions the commutation relations of $\mathfrak{psu}(1,1|2)_k$ are 
{\allowdisplaybreaks
\begin{subequations}\label{eq:psu112 commutation relations}
\begin{align} 
[J^3_m,J^3_n]&=-\tfrac{1}{2}km\delta_{m+n,0}\ ,  \label{eq:psu112 commutation relations a}\\
[J^3_m,J^\pm_n]&=\pm J^\pm_{m+n}\ , \label{eq:psu112 commutation relations b}\\
[J^+_m,J^-_n]&=km\delta_{m+n,0}-2J^3_{m+n}\ , \label{eq:psu112 commutation relations c}\\
[K^3_m,K^3_n]&=\tfrac{1}{2}km\delta_{m+n,0}\ , \label{eq:psu112 commutation relations d}\\
[K^3_m,K^\pm_n]&=\pm K^\pm_{m+n}\ , \label{eq:psu112 commutation relations e}\\
[K^+_m,K^-_n]&=km\delta_{m+n,0}+2K^3_{m+n}\ , \label{eq:psu112 commutation relations f}\\
[J^a_m,S^{\alpha\beta\gamma}_n]&=\tfrac{1}{2}c_a\, (\sigma^a)^\alpha_\mu \, S^{\mu\beta\gamma}_{m+n}\ , \label{eq:psu112 commutation relations g}\\
[K^a_m,S^{\alpha\beta\gamma}_n]&=\tfrac{1}{2} \, (\sigma^a)^\beta_\nu \, S^{\alpha\nu\gamma}_{m+n}\ , \label{eq:psu112 commutation relations h}\\
 \{S^{\alpha\beta\gamma}_m,S^{\mu\nu\rho}_n\}&=km \epsilon^{\alpha\mu}\epsilon^{\beta\nu}\epsilon^{\gamma\rho}\delta_{m+n,0}-\epsilon^{\beta\nu}\epsilon^{\gamma\rho} c_a \, (\sigma_a)^{\alpha\mu} J^a_{m+n}+\epsilon^{\alpha\mu}\epsilon^{\gamma\rho} (\sigma_a)^{\beta\nu} K^a_{m+n}\ , \label{eq:psu112 commutation relations i}
\end{align}
\end{subequations}}
where the $\sigma^a$ are Pauli matrices and $c_a= \pm 1$, see Appendix~A of \cite{Eberhardt:2018ouy} for more details.

The two descriptions\footnote{The last index can be raised and lowered via $\sqrt{2}S^{\alpha\beta+}=S^{\alpha\beta}_2$ and $-\sqrt{2}S^{\alpha\beta-}=S^{\alpha\beta}_{1}$.} are equivalent to one another, see also \cite{Seb}, and the translation between the different conventions is
{\allowdisplaybreaks
\begin{align}
J^{3}_{m}&=-\tfrac{i}{2}\left( K^{12}_{m}+K^{34}_{m} \right)\ ,& J^{\pm}_{m}&= \tfrac{1}{2}\left( K^{14}_{m}+K^{23}_{m} \pm iK^{13}_{m} \mp iK^{24}_{m} \right)\ , \\
K^{3}_{m}&=-\tfrac{i}{2}\left( K^{12}_{m}-K^{34}_{m} \right)\ ,& K^{\pm}_{m}&= \tfrac{1}{2}\left( \mp K^{14}_{m}\pm K^{23}_{m} + iK^{13}_{m} + iK^{24}_{m} \right) \ , \\
S^{++}_{\alpha,m}&=S^{1}_{\alpha,m}-iS^{2}_{\alpha,m}\ ,& S^{--}_{\alpha,m}&=S^{1}_{\alpha,m}+iS^{2}_{\alpha,m}\ , \\
S^{+-}_{\alpha,m}&=i S^{3}_{\alpha,m}+S^{4}_{\alpha,m}\ ,& S^{-+}_{\alpha,m}&=i S^{3}_{\alpha,m}-S^{4}_{\alpha,m}\ .
\label{4.1}
\end{align}}
Conversely, the inverse transformation is given by 
\begin{equation}
\begin{aligned}
J^{3}+K^{3}&=-iK^{12}\ ,\qquad & J^{3}-K^{3}&=-iK^{34}\ ,\\
J^{+}+J^{-}+K^{+}-K^{-}&=2K^{23}\ ,\qquad & J^{+}+J^{-}-K^{+}+K^{-}&=2K^{14}\ ,\\
J^{+}-J^{-}+K^{+}+K^{-}&=2iK^{13}\ ,\qquad & J^{+}-J^{-}-K^{+}-K^{-}&=-2iK^{24}\ ,
\label{4.3}
\end{aligned}
\end{equation}
while for the fermions we have 
\begin{equation}
\begin{aligned}
S^{++}_{\alpha}+S^{--}_{\alpha}&=2S^{1}_{\alpha}\ ,\qquad & S^{++}_{\alpha}-S^{--}_{\alpha}&=-2iS^{2}_{\alpha}\ ,\\
S^{+-}_{\alpha}+S^{-+}_{\alpha}&=2iS^{3}_{\alpha}\ ,\qquad & S^{+-}_{\alpha}-S^{-+}_{\alpha}&=2S^{4}_{\alpha}\ .
\label{4.5}
\end{aligned}
\end{equation}
Finally, the stress tensor of $\mathfrak{psu}(1,1|2)_k$ equals in either convention 
\begin{equation}
\begin{aligned}
T_{\mathfrak{psu}(1,1|2)}&=\frac{1}{8k}\epsilon^{abcd}\, : K^{ab}K^{cd} :  - \frac{1}{k} :(S^{a}_{1})(S^{a}_{2}) : \\
&= \frac{1}{k} :\Bigl( -J^{3}J^{3}+\frac{1}{2}(J^{+}J^{-}+J^{-}J^{+}) \Bigr) : +
\frac{1}{k} : \Bigl( K^{3}K^{3}+\frac{1}{2}(K^{+}K^{-}+K^{-}K^{+}) \Bigr) : \\
& \quad - \frac{1}{2k} :\left( S^{++}_{1}S^{--}_{2}+S^{--}_{1}S^{++}_{2}-S^{+-}_{1}S^{-+}_{2}-S^{-+}_{1}S^{+-}_{2} \right): \ ,
\label{A.16}
\end{aligned}
\end{equation}
where $: \cdots :$ denotes the normal ordering defined in eq.~(\ref{normalord}). Finally, we normal order the generators in (\ref{Rdef}) as 
\begin{equation} \label{A.17}
\begin{aligned}
(S^{a}_{\alpha}\partial S^{a}_{\alpha})_N &=\sum_{n \in \mathbb{Z}} -(n+1) :(S^{a}_{\alpha})_{N-n}(S^{a}_{\alpha})_{n}: \\
& =\sum_{n \geq 0} -(n+1) (S^{a}_{\alpha})_{N-n}(S^{a}_{\alpha})_{n}+ 
\sum_{n<0} (n+1) (S^{a}_{\alpha})_{n}(S^{a}_{\alpha})_{N-n} \ , 
\end{aligned}
\end{equation}
and
\begin{equation}\label{A.18}
\begin{aligned}
(K^{ab}S^{a}_{\alpha}S^{b}_{\alpha})_N & =\sum_{n, m \in \mathbb{Z}} :K^{ab}_{N-m-n}(S^{a}_{\alpha})_{m}(S^{b}_{\alpha})_{n}: \\ 
& =\sum_{\stackrel{{\tiny m \geq 0}}{N-m-n < 0}} -K^{ab}_{N-m-n}(S^{b}_{\alpha})_{n}(S^{a}_{\alpha})_{m} + \sum_{\stackrel{{\tiny m < 0}}{N-m-n < 0}} K^{ab}_{N-m-n}(S^{a}_{\alpha})_{m}(S^{b}_{\alpha})_{n}  \\
& \quad  +
\sum_{\stackrel{{\tiny m \geq 0}}{N-m-n \geq 0}} -(S^{b}_{\alpha})_{n}(S^{a}_{\alpha})_{m}K^{ab}_{N-m-n} + \sum_{\stackrel{{\tiny m < 0}}{N-m-n \geq 0}} (S^{a}_{\alpha})_{m}(S^{b}_{\alpha})_{n}K^{ab}_{N-m-n} \ .
\end{aligned}
\end{equation}

\subsection{The $\mathfrak{u}(1,1|2)_{k}$ algebra}\label{ulevk}

The other affine algebra that plays an important role in our analysis is $\mathfrak{u}(1,1|2)_{k}$. Its commutation relations are almost identical to those of $\mathfrak{psu}(1,1|2)_k$, see eq.~(\ref{eq:psu112 commutation relations}) above. Indeed, the eqs.~(\ref{eq:psu112 commutation relations a}) -- (\ref{eq:psu112 commutation relations h}) are the same, and instead of eq.~(\ref{eq:psu112 commutation relations i}) we now have 
\begin{align} 
 \{S^{\alpha\beta\gamma}_m,S^{\mu\nu\rho}_n\}&=km \epsilon^{\alpha\mu}\epsilon^{\beta\nu}\epsilon^{\gamma\rho}\delta_{m+n,0}-\epsilon^{\beta\nu}\epsilon^{\gamma\rho} c_a \, (\sigma_a)^{\alpha\mu} J^a_{m+n}+\epsilon^{\alpha\mu}\epsilon^{\gamma\rho} (\sigma_a)^{\beta\nu} K^a_{m+n} \nonumber \\
 & \quad + \epsilon^{\alpha\mu}\, \epsilon^{\beta\nu} \, \delta^{\gamma,-\rho} Z_{m+n} \ , \label{eq:u112 commutation relations i}
\end{align}
which differs from eq.~(\ref{eq:psu112 commutation relations i}) by the additional term proportional to $Z_{m+n}$. 
In addition, the algebra $\mathfrak{u}(1,1|2)_{k}$ contains the two generators $Y_m$ and $Z_n$ with 
\begin{subequations}\label{eq:u112 commutation relations1}
\begin{align}
[Z_m,J^a_n]&=0\ , & [Z_m,K^a_n]&=0\ , & [Z_m,S^{\alpha\beta\gamma}_n]&= 0 \ , \\
[Y_m,J^a_n]&=0\ , & [Y_m,K^a_n]&=0\ , & [Y_m,S^{\alpha\beta\gamma}_n]&=-\gamma S^{\alpha\beta\gamma}_{m+n}\ , \\
[Y_m,Y_n]&=0\ , & [Y_m,Z_n]&=-m \delta_{m+n,0}\ .
\end{align}
\end{subequations}

\section{Free field realisation of $\mathfrak{u}(1,1|2)_{1}$}\label{ffpsu1}

The free fields that enter in the free field realisation are four symplectic bosons, $\xi^\alpha$, $\eta^{\alpha}$, and four free fermions $\psi^\alpha$, $\chi^\alpha$, where $\alpha\in\{\pm\}$ and the OPEs are given as, see eq.~(\ref{freefields}),
\begin{equation}
\begin{aligned}
\xi^{+}(z)\eta^{-}(w) &\sim \dfrac{1}{z-w}\ ,& \qquad \xi^{-}(z)\eta^{+}(w) &\sim -\dfrac{1}{z-w}\ , \\ 
\psi^{+}(z)\chi^{-}(w) &\sim \dfrac{1}{z-w}\ ,& \qquad \psi^{-}(z)\chi^{+}(w) &\sim -\dfrac{1}{z-w}\ .
\label{freefields}
\end{aligned}
\end{equation}
In terms of these free fields, we can express the currents of $\mathfrak{u}(1,1|2)_1$ via 
{\allowdisplaybreaks\begin{equation}
\begin{aligned}
J^{3}&=-\tfrac{1}{2} \altcolon\left( \eta^{+}\xi^{-} + \eta^{-}\xi^{+}\right)\altcolon\ , &\quad J^{\pm}_{m}=& \altcolon\left( \eta^{\pm}\xi^{\pm}\right)\altcolon \ , \\
K^{3}&=-\tfrac{1}{2}\altcolon\left( \chi^{+}\psi^{-} + \chi^{-}\psi^{+}\right)\altcolon\ , &\quad K^{\pm}_{m}=&  \, \pm \altcolon\left( \chi^{\pm}\psi^{\pm}\right)\altcolon \ , \\
S^{\alpha \beta}_{2}&=\sqrt{2}\altcolon \xi^{\alpha}\chi^{\beta}\altcolon\ , &\quad S^{\alpha \beta}_{1}=&\, \sqrt{2} \altcolon\eta^{\alpha} \psi^{\beta}\altcolon\ , \\
U&=-\tfrac{1}{2}\altcolon\left( \eta^{+}\xi^{-} - \eta^{-}\xi^{+}\right)\altcolon\ , &\quad V=&-\tfrac{1}{2}\altcolon\left( \chi^{+}\psi^{-} - \chi^{-}\psi^{+}\right)\altcolon \ , 
\label{}
\end{aligned}
\end{equation}}
see also \cite{Eberhardt:2018ouy,DGGK}. Note that we could have equally normal ordered by $: \cdots :$ here, see eqs.~(\ref{CONCANO}) and (\ref{Rnormal}), since the two presecriptions are the same for the free fields. For future convenience we also define the two linear combinations 
\be
Z=U+V\ , \qquad Y=U-V \ . 
\ee
These fields generate indeed $\mathfrak{u}(1,1|2)_1$ since their OPEs equal 
{\allowdisplaybreaks \begin{equation}
\begin{aligned}
J^{3}(z)J^{3}(w)& \sim \dfrac{-1}{2(z-w)^{2}}\ ,& K^{3}(z)K^{3}(w)&\sim \dfrac{1}{2(z-w)^{2}} \ , \\
J^{3}(z)J^{\pm}(w)& \sim \dfrac{\pm J^{\pm}(w)}{z-w}\ ,& K^{3}(z)K^{\pm}(w)&\sim \dfrac{\pm K^{\pm}(w)}{z-w}\ , \\
J^{+}(z)J^{-}(w)&\sim \dfrac{-2J^{3}(w)}{z-w}+\dfrac{1}{(z-w)^{2}}\ ,& K^{+}(z)K^{-}(w)&\sim \dfrac{2K^{3}(w)}{z-w}+\dfrac{1}{(z-w)^{2}} \ , \\
J^{3}(z)S^{\pm \gamma}_{\alpha}(w)&\sim \pm \dfrac{1}{2}\dfrac{S^{\pm \gamma}_{\alpha}(w)}{z-w}\ ,& 
K^{3}(z)S^{\beta \pm}_{\alpha}(w)&\sim \pm \dfrac{1}{2}\dfrac{S^{\beta \pm}_{\alpha}(w)}{z-w} \ , \\
J^{\pm}(z)S^{\mp \gamma}_{\alpha}(w)&\sim \pm \dfrac{S^{\pm \gamma}_{\alpha}(w)}{z-w}\ ,& 
K^{\pm}(z)S^{\beta \mp}_{\alpha}(w)&\sim  \dfrac{S^{\beta \pm}_{\alpha}(w)}{z-w} \ , \\
S^{\pm \pm}_{\alpha}(z)S^{\pm \mp}_{\beta}(w)&\sim \pm \dfrac{2 \epsilon_{\alpha \beta}J^{\pm}(w)}{z-w}\ ,& S^{\pm \pm}_{\alpha}(z)S^{\mp \pm}_{\beta}(w)&\sim -  \dfrac{2 \epsilon_{\alpha \beta}K^{\pm}(w)}{z-w}\ , \\
Z(z)Y(w) &\sim -\dfrac{1}{(z-w)^2}\ ,& Y(z)S^{\alpha \beta}_{n}(w) &\sim \dfrac{(-1)^{n+1}S^{\alpha \beta}_{n}(w)}{z-w}\ ,
\label{4.11}
\end{aligned}
\end{equation}}
along with the OPEs of $SS$ that have a $Z$ term, 
\begin{equation} \label{eq:ss-z}
\begin{aligned}
S^{++}_{\alpha}(z)S^{--}_{\beta}(w)&\sim \dfrac{-2\epsilon_{\alpha \beta}(J^{3}-K^{3})(w)}{z-w}+\dfrac{2 \epsilon_{\alpha \beta}}{(z-w)^{2}}-\dfrac{2Z(w)|\epsilon_{\alpha\beta}|}{z-w}\ , \\
S^{+-}_{\alpha}(z)S^{-+}_{\beta}(w)&\sim \dfrac{2\epsilon_{\alpha \beta}(J^{3}+K^{3})(w)}{z-w}-\dfrac{2 \epsilon_{\alpha \beta}}{(z-w)^{2}}+\dfrac{2Z(w)|\epsilon_{\alpha\beta}|}{z-w} \ .
\end{aligned}
\end{equation}
Note that the difference to the $\mathfrak{psu}(1,1|2)_1$ current OPEs are precisely these $Z$-terms, and thus we need to `gauge' by the $Z$ field in order to reduce $\mathfrak{u}(1,1|2)_1$ to $\mathfrak{psu}(1,1|2)_1$.

\section{Normal ordering conventions}\label{NO}

Normal ordering for products of two fields is relatively straightforward, see e.g.\ eq.~(\ref{normalord}), but whenever more fields are involved, as is for example the case in the definition of (\ref{S4def}) and (\ref{Rdef}), 
there are ambiguities, and it will be important for us to fix them.  In the following we will explain, following \cite{Pol}, a definite prescription for how to normal order an arbitrary number of fields. We will denote the corresponding normal ordering by $\altcolon\cdot \altcolon$ and sometimes refer to it as `conformal normal ordering'. For the free symplectic bosons and fermions fields, this normal ordering agrees with (\ref{CONCANO}) and (\ref{Rnormal}) when restricted to two fields. However, we will also apply it in general, and this is how we define the triple products in (\ref{S4def}) and (\ref{Rdef}).

In the following we shall first explain the prescription for a free boson field; at the end of this appendix  we shall  also explain how it works for the $bc$ system, as well as a chiral boson, see eqs.~(\ref{bc}) and (\ref{chiralb}) below.  The OPE of the free boson field $X(z)$ is 
\be\label{XOPE}
\Bigl(\, \overbracket{X(z)X(w)}\, \Bigr) =-\frac{\alpha'}{2}\ln(|z-w|^{2})\ .
\ee
We define the normal ordering by 
\be
\altcolon X(z)X(w)\altcolon\, =X(z)X(w)+\dfrac{\alpha'}{2}\ln(|z-w|^{2})\ ,
\ee
where the first term on the right hand side is radially ordered, i.e.\ the field with bigger modulus stands to the left of the one with smaller modulus.  Our normal ordering is therefore equivalent to acting by a differential operator 
\begin{equation}
\begin{aligned}
\altcolon X(z)X(w)\altcolon &=X(z)X(w)+\dfrac{\alpha'}{2}\ln(|z-w|^{2})\\
&=\exp\left(\int d^{2}x\int d^{2}y\left(\dfrac{\alpha'}{4}\ln(|x-y|^{2})\right)\dfrac{\delta}{\delta X(x)}\dfrac{\delta}{\delta X(y)}\right)X(z)X(w)\ .
\end{aligned}
\end{equation}
We can invert this relation to give
\begin{equation}
\begin{aligned}
X(z)X(w)&=\altcolon\left[\exp\left(\int d^{2}x\int d^{2}y\left(-\dfrac{\alpha'}{4}\ln(|x-y|^{2})\right)\dfrac{\delta}{\delta X(x)}\dfrac{\delta}{\delta X(y)}\right)X(z)X(w)\right]\altcolon \ , 
\label{A.2}
\end{aligned}
\end{equation}
since the differential operators in the exponent commute with one another. 
The normal ordering of three or more fields can then be defined similarly, i.e.
\begin{equation}
\begin{aligned}
\altcolon X(z)X(w)X(t)\altcolon:=\exp\left(\int d^{2}x\int d^{2}y\left(\dfrac{\alpha'}{4}\ln(|x-y|^{2})\right)\dfrac{\delta}{\delta X(x)}\dfrac{\delta}{\delta X(y)}\right)X(z)X(w)X(t) \ , 
\end{aligned}
\end{equation}
where the fields on the right hand side are again radially ordered. We can also invert this relation as in eq.~(\ref{A.2}).

It should now be clear how this can also be generalised to obtain the  radial ordering of two normal ordered operators 
\be 
\altcolon\mathcal{F}(z)\altcolon\, \altcolon\mathcal{G}(w)\altcolon\ ,
\ee
where $\mathcal{F},\mathcal{G}$ are functions of the field $X$. Using the above formula for normal ordering, we have
\begin{equation}
\begin{aligned}
\altcolon\!\mathcal{F}(z)\!\altcolon\, \altcolon\!\mathcal{G}(w)\!\altcolon&=\, \altcolon\!\!\left[\exp\left(\int d^{2}x\int d^{2}y\left(-\dfrac{\alpha'}{4}\ln(|x-y|^{2})\right)\dfrac{\delta}{\delta_{\mathcal{F},\mathcal{G}} X(x)}\dfrac{\delta}{\delta_{\mathcal{F},\mathcal{G}} X(y)}\right)\altcolon\!\mathcal{F}\!\altcolon\, \altcolon\!\mathcal{G}\!\altcolon\right]\!\!\altcolon\\
&=\, \altcolon\!\!\left[ \exp\left(\int d^{2}x\int d^{2}y\left(-\dfrac{\alpha'}{4}\ln(|x-y|^{2})\right)\dfrac{\delta}{\delta_{\mathcal{F},\mathcal{G}} X(x)}\dfrac{\delta}{\delta_{\mathcal{F},\mathcal{G}} X(y)}\right)\times\right.\\
&~~~~~~\left.\exp\left(\int d^{2}x\int d^{2}y\left(\dfrac{\alpha'}{4}\ln(|x-y|^{2})\right)\dfrac{\delta}{\delta_{\mathcal{F}} X(x)}\dfrac{\delta}{\delta_{\mathcal{F}} X(y)}\right)\times\right.\\
&~~~~~~\left.\exp\left(\int d^{2}x\int d^{2}y\left(\dfrac{\alpha'}{4}\ln(|x-y|^{2})\right)\dfrac{\delta}{\delta_{\mathcal{G}} X(x)}\dfrac{\delta}{\delta_{\mathcal{G}} X(y)}\right)\right]\mathcal{F}\mathcal{G}\altcolon\ ,
\end{aligned}
\end{equation}
where 
\be
\frac{\delta}{\delta_{\mathcal{F},\mathcal{G}} X(x)}=\frac{\delta}{\delta_{\mathcal{F}} X(x)}+\frac{\delta}{\delta_{\mathcal{G}} X(x)} \ , 
\ee
and $\frac{\delta}{\delta_{\mathcal{F}} X(x)}$ is the differentiation with respect to $X(x)$ of $\mathcal{F}(X)$, and similarly for $\mathcal{G}$. Altogether, this therefore leads to  
\begin{equation}
\begin{aligned}
&\altcolon\!\mathcal{F}(z)\!\altcolon\, \altcolon\!\mathcal{G}(w)\!\altcolon\, =\, \altcolon\left[\exp\left(\int d^{2}x\int d^{2}y\left(-\dfrac{\alpha'}{2}\ln(|x-y|^{2})\right)\dfrac{\delta}{\delta_{\mathcal{F}} X(x)}\dfrac{\delta}{\delta_{\mathcal{G}} X(y)}\right)\mathcal{F}\mathcal{G}\right]\altcolon.
\label{7.5}
\end{aligned}
\end{equation}
We can, in particular, use this formula to compute radial orderings of products of exponentials. Since 
$\exp(imX(z))\exp(inX(w))$ is an eigenfunction of 
\be
\int d^{2}x\int d^{2}y\left(-\dfrac{\alpha'}{2}\ln(|x-y|^{2})\right)\dfrac{\delta}{\delta_{\mathcal{F}} X(x)}\dfrac{\delta}{\delta_{\mathcal{G}} X(y)} \ , 
\ee
i.e.,
\begin{equation}
\begin{aligned}
\int d^{2}x\int d^{2}y\left(-\dfrac{\alpha'}{2}\ln(|x-y|^{2})\right)\dfrac{\delta}{\delta_{\mathcal{F}} X(x)}\dfrac{\delta}{\delta_{\mathcal{G}} X(y)}& \, e^{i mX}(z) \, e^{i nX}(w)  \\
= \frac{\alpha'mn\ln(|z-w|^{2})}{2}&\, e^{i mX}(z) \, e^{i nX}(w) \ , 
\label{}
\end{aligned}
\end{equation}
eq.~(\ref{7.5}) becomes
\begin{equation}
\begin{aligned}
\altcolon \! e^{i mX}(z) \!\altcolon\, \altcolon \!e^{i nX}(w) \!\altcolon\, &=\, \altcolon \!\!\left[e^{\frac{\alpha'mn\ln(|z-w|^{2})}{2}}\, e^{i mX}(z)e^{i nX}(w) \right] \!\!\altcolon\\
&=|z-w|^{\alpha'mn}\, \altcolon\! e^{i mX}(z) e^{i nX}(w) \!\altcolon\ .
\label{}
\end{aligned}
\end{equation}


The generalisation from the free boson to a $(b,c)$ system is straightforward. For a $(b,c)$ system we have instead of (\ref{XOPE}),
\be\label{bc}
\Bigl(\, \overbracket{b(z)c(w)}\, \Bigr)=\dfrac{1}{z-w}\ ,
\ee
and hence 
\begin{equation}
\begin{aligned}
&\altcolon\!\mathcal{F}(z)\!\altcolon\, \altcolon\!\mathcal{G}(w)\!\altcolon\, =\, \altcolon\!\!\left[\exp\left(\int d^{2}x\int d^{2}y\left(\dfrac{1}{x-y}\right)\left(\dfrac{\delta}{\delta_{\mathcal{G}} c(y)}\dfrac{\delta}{\delta_{\mathcal{F}} b(x)}+\dfrac{\delta}{\delta_{\mathcal{F}} c(y)}\dfrac{\delta}{\delta_{\mathcal{G}} b(x)}\right)\right)\mathcal{F}\mathcal{G}\right]\!\!\altcolon.\\
\label{7.9}
\end{aligned}
\end{equation}
Finally, for a chiral boson with OPE
\be\label{chiralb}
\Bigl(\, \overbracket{\rho(z)\rho(w)}\, \Bigr)=-\ln(z-w)\ ,
\ee
we get 
\begin{equation}
\begin{aligned}
&\altcolon\mathcal{F}(z)\altcolon\altcolon\mathcal{G}(w)\altcolon=\altcolon\left[\exp\left(\int d^{2}x\int d^{2}y\left(-\ln(x-y)\right)\dfrac{\delta}{\delta_{\mathcal{F}} \rho(x)}\dfrac{\delta}{\delta_{\mathcal{G}} \rho(y)}\right)\mathcal{F}\mathcal{G}\right]\altcolon.
\label{7.10}
\end{aligned}
\end{equation}

\section{Checking the ${\cal N}=2$ algebra}

In this appendix we give some details concerning the calculation of the OPEs of the ${\cal N}=2$ generators of eq.~(\ref{N2gencur}). 

\subsection{Calculation of the $G^{+}G^{+}$ OPE}\label{GG}

Let us start with the $G^{+}G^{+}$ OPE. We shall concentrate on the computation of the $Q_{0}Q_{0}$ OPE, which contains the contribution, see eq.~(\ref{GPdef}) --- the remaining terms will be discussed below
\begin{equation}
\begin{aligned}
&\altcolon e^{i\sigma(z)} \Bigl( - \frac{1}{2} (\partial_{z} (\rho + i \sigma))^{2}+ \frac{1}{2} \partial^2_{z} (\rho + i \sigma) \Bigr)\altcolon\, \altcolon e^{i\sigma(w)} \Bigl( - \frac{1}{2} (\partial_{w} (\rho + i \sigma))^{2}+ \frac{1}{2} \partial^2_{w} (\rho + i \sigma) \Bigr)\altcolon\\
&\quad =\, \altcolon\left[\exp\left(\int d^{2}x\int d^{2}y\left(-\ln(x-y)\right)\left(\dfrac{\delta}{\delta_{\mathcal{F}} \rho(x)}\dfrac{\delta}{\delta_{\mathcal{G}} \rho(y)}+\dfrac{\delta}{\delta_{\mathcal{F}} \sigma(x)}\dfrac{\delta}{\delta_{\mathcal{G}} \sigma(y)}\right)\right)\mathcal{F}\mathcal{G}\right]\altcolon.
\label{7.11}
\end{aligned}
\end{equation}
Here we are using the conformal normal ordering convention, which was explained in Appendix~\ref{NO}.
By the usual chain rule argument we see that
\begin{equation}
\begin{aligned}
\dfrac{\delta}{\delta_{\mathcal{F}} \rho(x)}&=\int d^{2}y\dfrac{\delta}{\delta_{\mathcal{F}} (\rho(y)+i\sigma(y))} \dfrac{\delta_{\mathcal{F}} (\rho(y)+i\sigma(y))}{\delta_{\mathcal{F}} \rho(x)}=\int d^{2}y\dfrac{\delta}{\delta_{\mathcal{F}} (\rho(y)+i\sigma(y))} \delta^{2}(x-y)\\
&=\dfrac{\delta}{\delta_{\mathcal{F}} (\rho(x)+i\sigma(x))}\ ,
\end{aligned}
\end{equation}
and similarly
\be
\dfrac{\delta}{\delta_{\mathcal{F}} \sigma(x)}=i\dfrac{\delta}{\delta_{\mathcal{F}} (\rho(x)+i\sigma(x))}+\dfrac{\delta}{\delta_{\mathcal{F}} \sigma(x)}\ .
\ee
Thus the differential operator in \eqref{7.11} simplifies to,
\begin{equation}
\begin{aligned}
&\left[\exp\left(\int d^{2}x\int d^{2}y\left(-\ln(x-y)\right)\times \right. \right.\\
&\left.\left.\left(\dfrac{\delta}{\delta_{\mathcal{F}} \sigma(x)}\dfrac{\delta}{\delta_{\mathcal{G}} \sigma(y)}+i\dfrac{\delta}{\delta_{\mathcal{F}} (\rho(x)+i\sigma(x))}\dfrac{\delta}{\delta_{\mathcal{G}} \sigma(y)}+i\dfrac{\delta}{\delta_{\mathcal{F}} \sigma(x)}\dfrac{\delta}{\delta_{\mathcal{G}} (\rho(y)+i\sigma(y))}\right)\right)\right].
\label{7.12}
\end{aligned}
\end{equation}
The first term in \eqref{7.12} tells us to contract the exponentials $e^{i\sigma(z)}\, e^{i\sigma(w)}$, and this simply gives the factor 
\be\label{eeterm}
e^{i\sigma(z)}\, e^{i\sigma(w)} \sim (z-w) \ . 
\ee
The remaining two terms in \eqref{7.12} lead to contractions between the exponential term, and the derivative terms. They can be evaluated straightforwardly, and the result is 
\begin{equation}
\begin{aligned}
&\altcolon e^{i\sigma(z)} \Bigl( - \frac{1}{2} (\partial_{z} (\rho + i \sigma))^{2} + \frac{1}{2} \partial^2_{z} (\rho + i \sigma) \Bigr)\altcolon \altcolon e^{i\sigma(w)} \Bigl( - \frac{1}{2} (\partial_{w} (\rho + i \sigma))^{2} + \frac{1}{2} \partial^2_{w} (\rho + i \sigma) \Bigr)\altcolon\\
&=\, \altcolon\dfrac{e^{i\sigma(z)}}{4}(z-w)\left( (\partial_{z} (\rho + i \sigma))^{2} - \partial^2_{z} (\rho + i \sigma)+\dfrac{2}{(z-w)^{2}}+\dfrac{2\partial_{z}(\rho+i\sigma)}{z-w} \right) \times\\
&~~~~\left( (\partial_{w} (\rho + i \sigma))^{2} - \partial^2_{w} (\rho + i \sigma)+\dfrac{2}{(z-w)^{2}}-\dfrac{2\partial_{w}(\rho+i\sigma)}{z-w} \right)e^{i\sigma(w)}\altcolon\\
&=\dfrac{\altcolon e^{i\sigma(z)}e^{i\sigma(w)}\altcolon}{(z-w)^{3}}\ .
\label{7.13}
\end{aligned}
\end{equation}
The remaining contributions of the $Q_0 Q_0$ OPE involve $T_{\rm int}$. 
The contraction of the $T_{\rm int}$ term with the derivatives leads to 
\be\label{D.6}
e^{i\sigma(z)}T_{\rm int}(z) \cdot e^{i\sigma(w)} \Bigl( - \frac{1}{2} \partial_{w} (\rho + i \sigma) \partial_{w} (\rho + i \sigma) + \frac{1}{2} \partial^2_{w} (\rho + i \sigma) \Bigr) \sim 
\frac{-\, T_{\rm int}(w)\, e^{2i\sigma(w)}}{z-w} \ ,
\ee
as can be found by similar techniques. Finally, we find for the $T_{\rm int} \cdot T_{\rm int}$ term 
\be\label{D.7}
e^{i\sigma(z)}T_{\rm int}(z)\, e^{i\sigma(w)}T_{\rm int}(w) \, \sim \, \altcolon e^{i\sigma(z)}e^{i\sigma(w)}\altcolon \left( \frac{-1}{(z-w)^{3}}+\dfrac{2T_{\rm int}(w)}{z-w} \right)\ .
\ee
Thus the sum of eqs.~(\ref{7.13}), twice (\ref{D.6}) --- there are two such terms, and they contribute equally since the relevant operators are fermionic --- and (\ref{D.7})  cancel. Note that the leading term in the last OPE comes from the central charge contribution of the $T_{\rm int} \cdot T_{\rm int}$ OPE, using that $c=-2$, as well as eq.~(\ref{eeterm}).
%
The other OPEs can be checked similarly. 


\subsection{The fermionic symmetry generator}\label{app:symgen}

In this appendix we explain that the simple pole in the $Q_1 \cdot S^{a}_{2}$ OPE cancels against that of the $Q_0 \cdot e^{-\rho-i\sigma}\, S^a_1$ OPE, see the discussion below eq.~(\ref{fermsymmod}) in the main part of the paper. Let us begin with the latter OPE, 
\be
e^{-\rho(z)-i\sigma(z)}S^{l}_{1}(z)\, e^{i\sigma(w)}\Bigl(T_{\rm int}- \frac{1}{2} \partial (\rho + i \sigma) \partial (\rho + i \sigma) + \frac{1}{2} \partial^2 (\rho + i \sigma)\Bigr)(w) \ . 
\ee
The contraction of the ghost terms can be done using the same methods as in Appendix~\ref{GG}, and  it leads to a factor of 
\begin{align}
& e^{-\rho(z)-i\sigma(z)} S^{l}_{1}(z) \, e^{i\sigma(w)}\Bigl(- \frac{1}{2} \partial (\rho + i \sigma) \partial (\rho + i \sigma) + \frac{1}{2} \partial^2 (\rho + i \sigma)\Bigr)(w) \\
& \qquad \sim -\frac{e^{-\rho(z)-i\sigma(z)+i\sigma(w)}}{(z-w)} \, S^{l}_{1}(z) \, \Bigl(- \frac{1}{2} \partial (\rho + i \sigma) \partial (\rho + i \sigma) + \frac{1}{2} \partial^2 (\rho + i \sigma)\Bigr)(w) \ . 
\label{1}
\end{align}
On the other hand, the contribution coming from the OPE of $S^{l}_{1}(z)\,$ with $T_{\rm int}(w)$ gives rise to 
\be\label{2}
e^{-\rho(z)-i\sigma(z)} S^{l}_{1}(z) \, e^{i\sigma(w)} T_{\rm int}(w) \sim 
-\frac{e^{-\rho(z)-i\sigma(z)+i\sigma(w)}}{(z-w)} \, \Bigl[ \frac{S^{l}_{1}(w)}{(z-w)^{2}}+:S^{l}_{1}(w)T_{\rm int}(w): \Bigr] \ . 
\ee
Expanding out the exponential in  (\ref{2}), and concentrating on the simple pole --- this is the only contribution that survives after taking the contour integral --- the first term in the bracket leads to contributions that cancel exactly the simple pole in (\ref{1}). 
On the other hand, the second term in the bracket of (\ref{2}) cancels against part of the simple pole of $S^{l}_{2}(z)Q_{1}(w)$, that can be calculated directly. The terms that remain from the simple pole of $\tilde{S}^{l}_{2}(z)G^+(w)$ are proportional to $e^{-\rho} \Phi$,\footnote{Recall that $e^{-\rho}$ has conformal dimension $h=-2$.}  where
\begin{equation}
\begin{aligned}
\Phi = -\frac{1}{8}\left( \epsilon^{abcd}K^{ab}_{-1}K^{cd}_{-1}(S^{l}_{1})_{-1}+4\epsilon^{lamn}K^{ab}_{-1}K^{mn}_{-1}(S^{b}_{1})_{-1}-4\epsilon^{labd}K^{cb}_{-1}K^{cd}_{-1}(S^{a}_{1})_{-1} \right)\ket{0} \ ,
\label{deltaG}
\end{aligned}
\end{equation}
and we have used that 
\be
\epsilon^{lamn}K^{mn}_{-2}=\epsilon^{labd}K^{cb}_{-1}K^{cd}_{-1} \ , 
\ee
as follows from the commutation relation (\ref{3.19}). By contracting the last two terms in  \eqref{deltaG} with $\epsilon^{lijk}$, we find 
\begin{equation}
\begin{aligned}
& \epsilon^{lijk} \bigl( \epsilon^{lamn}K^{ab}_{-1}K^{mn}_{-1}(S^{b}_{1})_{-1}-\epsilon^{labd}K^{cb}_{-1}K^{cd}_{-1}(S^{a}_{1})_{-1} \Bigr) \\
& \qquad\quad = 2\left[ K^{ib}_{-1}K^{jk}_{-1}(S^{b}_{1})_{-1}+K^{kb}_{-1}K^{ij}_{-1}(S^{b}_{1})_{-1}+K^{jb}_{-1}K^{ki}_{-1}(S^{b}_{1})_{-1} \right]\\
& \qquad\qquad - \left[ K^{cj}_{-1}K^{ck}_{-1}(S^{i}_{1})_{-1}+K^{ci}_{-1}K^{cj}_{-1}(S^{k}_{1})_{-1}+K^{ck}_{-1}K^{ci}_{-1}(S^{j}_{1})_{-1}\right]\\
& \qquad\qquad + \left[K^{cj}_{-1}K^{ci}_{-1}(S^{k}_{1})_{-1}+K^{ck}_{-1}K^{cj}_{-1}(S^{i}_{1})_{-1}+K^{ci}_{-1}K^{ck}_{-1}(S^{j}_{1})_{-1} \right]\ .
\label{b5}
\end{aligned}
\end{equation}
By direct evaluation we see that the terms that contain $(S^{g}_{1})_{-1}$ where $g$ is either $g=i$, or $g=j$, or $g=k$ vanish, and hence $g$ has to take the remaining value $g \in \{1,2,3,4\}\textbackslash\{i,j,k\}$. Thus, 
eq.~\eqref{b5} becomes
\begin{align}
& \epsilon^{lijk} \bigl( \epsilon^{lamn}K^{ab}_{-1}K^{mn}_{-1}(S^{b}_{1})_{-1}-\epsilon^{labd}K^{cb}_{-1}K^{cd}_{-1}(S^{a}_{1})_{-1} \Bigr) \\
& \qquad = 
2\left( K^{il}_{-1}K^{jk}_{-1}(S^{l}_{1})_{-1}+K^{kl}_{-1}K^{ij}_{-1}(S^{l}_{1})_{-1}+K^{jl}_{-1}K^{ki}_{-1}(S^{l}_{1})_{-1} \right) \ ,
\end{align}
where $l$ is a free label. Contracting with $\epsilon^{pijk}$ we get 
\begin{equation}
\epsilon^{pamn}K^{ab}_{-1}K^{mn}_{-1}(S^{b}_{1})_{-1}-\epsilon^{pabd}K^{cb}_{-1}K^{cd}_{-1}(S^{a}_{1})_{-1}=\epsilon^{pijk}K^{ip}_{-1}K^{jk}_{-1}(S^{p}_{1})_{-1} \ , 
\label{coolrel}
\end{equation}
where now the free index is $p$. Finally, we can rewrite for fixed $p$ the product of the two $K$ operators on the right as 
\begin{equation}
\epsilon^{pijk}K^{ip}_{-1}K^{jk}_{-1}=-\tfrac{1}{4}\epsilon^{abcd}K^{ab}_{-1}K^{cd}_{-1}\ .
\label{supercoolrel}
\end{equation}
(Here the factor of $\tfrac{1}{4}$ arises because $p$ is not summed over the $4$ values, and the minus sign arises because we exchanged the order of the indices in the totally anti-symmetric tensor.) 
Then, combining \eqref{coolrel} and \eqref{supercoolrel}, it follows that $\Phi$ in eq.~\eqref{deltaG} indeed vanishes.

\subsection{Calculation of the $R$ in terms of the free fields}\label{Rterm}

In this appendix we explain the calculation of $R$, see eq.~(\ref{Rdef}) for $k=1$, in terms of the free fields, see eq.~(\ref{Q1free}). In terms of $\mathfrak{psu}(1,1|2)_1$ generators the state corresponding to $R$ equals in our conventions 
\begin{equation}
\begin{aligned}
R=&-\frac{1}{2}\Bigl[  -(J^{3}_{-1}+K^{3}_{-1})\, (S^{--}_{1})_{-1} (S^{++}_{1})_{-1}+
 (J^{3}_{-1}-K^{3}_{-1})\, (S^{-+}_{1})_{-1} (S^{+-}_{1})_{-1}   \\
&  \qquad -J^{+}_{-1}(S^{--}_{1})_{-1} (S^{-+}_{1})_{-1}+ J^{-}_{-1} (S^{++}_{1})_{-1}(S^{+-}_{1})_{-1} +4(S\partial S)_{-3} \Bigr] \ket{0}\ ,
\label{R0}
\end{aligned}
\end{equation}
where we have already used that $(S^{\alpha \beta}_{1})_{-1} (S^{\gamma \beta}_{1})_{-1} \ket{0}=0$ in the free field realisation. Plugging in the free field realisation for these generators, we find for the terms in the first line
\begin{align}
(S^{--}_{1})_{-1} (S^{++}_{1})_{-1} \ket{0} & =2\, \eta^{-}_{-1/2}\psi^{-}_{-1/2}\eta^{+}_{-1/2}\psi^{+}_{-1/2}\ket{0}\\
(S^{-+}_{1})_{-1} (S^{+-}_{1})_{-1} \ket{0} & =2\, \eta^{-}_{-1/2}\psi^{+}_{-1/2}\eta^{+}_{-1/2}\psi^{-}_{-1/2}\ket{0} = - (S^{--}_{1})_{-1} (S^{++}_{1})_{-1} \ket{0} \ . 
\end{align}
Thus only the $J^3_{-1}$ descendant of the first line survives, and it leads to 
\begin{align}
- J^{3}_{-1} (S^{--}_{1})_{-1} (S^{++}_{1})_{-1} \ket{0} & =  \Bigl[-  \bigl(\eta^+_{-1/2} \xi^-_{-1/2} + \eta^-_{-1/2} \xi^+_{-1/2} \bigr)\,  \eta^{-}_{-1/2}\eta^{+}_{-1/2} \\
&  \qquad 
+  \eta^{-}_{-1/2}\eta^{+}_{-3/2} 
-  \eta^{-}_{-3/2}\eta^{+}_{-1/2} \Bigr]  \psi^{+}_{-1/2} \psi^{-}_{-1/2} \ket{0} \ .
\end{align}
For the terms containing boson generators in the second line we find similarly
\begin{align}
-J^{+}_{-1}S^{--}_{1}S^{-+}_{1}\ket{0}&= -2(\eta^{+}_{-1/2}\xi^{+}_{-1/2}+\eta^{+}_{-3/2}\xi^{+}_{1/2})\, \eta^{-}_{-1/2}\psi^{-}_{-1/2}\eta^{-}_{-1/2}\psi^{+}_{-1/2}\ket{0}\\
&= 2\Bigl[  \eta^{+}_{-1/2}\xi^{+}_{-1/2}\eta^{-}_{-1/2}\eta^{-}_{-1/2}+ 2\eta^{+}_{-3/2}\eta^{-}_{-1/2}\Bigr] \psi^{+}_{-1/2} \psi^{-}_{-1/2} \ket{0}  \ , 
\end{align}
and 
\begin{align}
J^{-}_{-1}S^{++}_{1}S^{+-}_{1}\ket{0}&= 2(\eta^{-}_{-1/2}\xi^{-}_{-1/2}+\eta^{-}_{-3/2}\xi^{-}_{1/2})\, \eta^{+}_{-1/2}\psi^{+}_{-1/2}\eta^{+}_{-1/2}\psi^{-}_{-1/2}\ket{0}\\
&=  2\Bigl[  \eta^{-}_{-1/2}\xi^{-}_{-1/2}\eta^{+}_{-1/2}\eta^{+}_{-1/2} - 2 \eta^{-}_{-3/2}\eta^{+}_{-1/2}\Bigr] \psi^{+}_{-1/2} \psi^{-}_{-1/2} \ket{0}  \ .
\end{align}
Combining all these terms that involve bosonic generators, i.e.\ the $K^{ab}S^{a}_{1}S^{b}_{1}$ term in eq.\ (\ref{Rdef}), we thus obtain 
\begin{equation}
\begin{aligned}
K^{ab}S^{a}_{1}S^{b}_{1} =\Bigl[-6\, \psi^{+}\psi^{-}(\eta^{+}\partial \eta^{-}-\eta^{-} \partial \eta^{+})\Bigr] \ .\label{KSSfinal}
\end{aligned}
\end{equation}
On the other hand, the $4S^{a}_{1} \partial S^{a}_{1}$ term eq.~(\ref{Rdef}), i.e.\ the $4S \partial S$ term in \eqref{R0}, equals 
\begin{equation}
\begin{aligned}
4S^{a}_{1} \partial S^{a}_{1}  =\Bigl[8\psi^{+}\psi^{-}(\eta^{+}\partial \eta^{-}-\eta^{-} \partial \eta^{+})\Bigr]\ . 
\label{}
\end{aligned}
\end{equation}
Altogether we therefore find for $R$ at $k=1$, see eq.~(\ref{Rdef}) 
\be
R = -\frac{1}{2}\Bigl( K^{ab}S^{a}_{1}S^{b}_{1}+4S^{a}_{1} \partial S^{a}_{1} \Bigr) = - 
\psi^{+}\psi^{-}\, (\eta^{+}\partial \eta^{-}-\eta^{-} \partial \eta^{+}) \ ,
\ee
thus giving us the first equation in eq.~(\ref{Q1free}).

\end{document}